\documentclass[iop,appendixfloats]{emulateapj}
\usepackage{natbib}
\usepackage{amsmath}
\usepackage{xspace}
\usepackage{enumitem}

\newcommand{\hst}{{\it HST}\xspace}
\newcommand{\Z}{{\rm \lbrack Z/H\rbrack}\xspace}

\def\LKsun{\hbox{$\thinspace L_{K,\odot}$} }

\shorttitle{The stellar halo of NGC~3115}
\shortauthors{Peacock et al.}

\begin{document}

\title{Detection of a distinct metal-poor stellar halo in the early-type galaxy NGC~3115$^{\dag}$}

\author{Mark B. Peacock$^{1}$, Jay Strader$^{1}$, Aaron J. Romanowsky$^{2,3}$, Jean P. Brodie$^{3}$}
\affil{$^{1}$Department of Physics and Astronomy, Michigan State University, East Lansing, MI 48824, USA mpeacock@msu.edu}
\affil{$^{2}$Department of Physics and Astronomy, San Jos\'e State University, San Jose, CA 95192, USA}
\affil{$^{3}$University of California Observatories/Lick Observatory, Santa Cruz, CA 95064, USA}
\email{
$^{\dag}$ Based on observations made with the NASA/ESA Hubble Space Telescope, obtained at the Space Telescope Science Institute, which is operated by the Association of Universities for Research in Astronomy, Inc., under NASA contract NAS 5-26555. These observations are associated with program \#13048. 
}

\begin{abstract}
\label{sec:abstract}

We present the resolved stellar populations in the inner and outer halo of the nearby lenticular galaxy NGC~3115. Using deep \hst observations, we analyze stars two magnitudes fainter than the tip of the red giant branch (TRGB). We study three fields along the minor axis of this galaxy, 19, 37 and 54~kpc from its center -- corresponding to 7, 14, 21 effective radii ($r_{e}$). Even at these large galactocentric distances, all of the fields are dominated by a relatively enriched population, with the main peak in the metallicity distribution decreasing with radius from $\Z \sim -0.5$ to $-0.65$. The fraction of metal-poor stars ($\Z < -0.95$) increases from $17\%$, at $16-37$~kpc, to $28\%$, at $\sim$54~kpc. We observe a distinct low metallicity population (peaked at $\Z \sim -1.3$ and with total mass $2\times10^{10}M_{\odot} \sim 14\%$ of the galaxy's stellar mass) and argue that this represents the detection of an underlying low metallicity stellar halo. Such halos are generally predicted by galaxy formation theories and have been observed in several late type galaxies including the Milky Way and M31. The metallicity and spatial distribution of the stellar halo of NGC~3115 are consistent with the galaxy's globular cluster system, which has a similar low metallicity population that becomes dominant at these large radii. This finding supports the use of globular clusters as bright chemo-dynamical tracers of galaxy halos. These data also allow us to make a precise measurement of the magnitude of the TRGB, from which we derive a distance modulus of NGC~3115 of $30.05\pm 0.05\pm 0.10_{sys}$ ($10.2\pm0.2\pm0.5_{sys}$~Mpc). 

\end{abstract}


\section{Introduction}
\label{sec:intro}

The stellar halos of galaxies are relics of their past -- preserved by their low densities and long relaxation times. As such, they provide key clues to their assembly history \citep[][]{Searle78}. The presence of extended low metallicity stellar halos around all massive galaxies is a central prediction of hierarchical galaxy formation \citep[e.g.][]{Bullock05, Cooper10, Zolotov10, Cooper13}. These halos arise naturally through the accretion and stripping of lower mass galaxies and their entourages of globular clusters. The mass, metallicity and density profile of a galaxy's halo is related to the number, size and epoch of such interactions. The halo of a galaxy is, therefore, a prime location for understanding its past, and for constraining galaxy formation in general.

Beyond the Local Group of galaxies, most of our knowledge of galaxy halos comes from surface brightness photometry of the integrated stellar emission \citep[e.g.][]{Martinez-Delgado10,LaBarbera12,Mihos13,vanDokkum14,DSouza14}. For example, \citet{Mihos13} utilized such surface brightness photometry to probe the halo of NGC~4472 out to 7 effective radii ($r_{e}$), finding a color gradient that may be the result of the increasing contribution with radius of a low metallicity stellar population (${\rm [Fe/H]} < -1.0$). However, most studies of galaxies using this method are limited to a few $r_{e}$ and hence fail to probe the proposed accretion dominated outer halo. This is because the extremely low surface brightness of the outer halos of galaxies limits the radii that can be observed with most telescopes. This work also requires extremely accurate sky subtraction and flatfielding, making interpretation of observations challenging. In addition to these observational difficulties, integrated photometry can only determine the average properties of the stellar halo. Resolving the stellar populations of galaxy halos offers significantly more information, but is necessarily limited to nearby galaxies. 

The best studied galaxy halo remains the Milky Way's, which is observed to have an old, smooth and metal-poor stellar population \citep[with mean \lbrack Fe/H\rbrack~=~-1.6 in the inner halo,][]{Carollo10}. Resolved photometry of the stellar populations of the local and similar spiral galaxies NGC~891, M~81 and M~31 have shown that they also have a distinct metal-poor halo \citep{Ibata09, Monachesi13, Ibata14}. In M31, this metal poor halo has also been confirmed via spectroscopy of its red giant branch stars \citep{Gilbert14}. However, the well studied stellar halo of M~31 differs from the Milky Way in two important ways: first, in addition to an underlying old metal-poor population, there is a substantial population of higher metallicity stars; second, a larger fraction of M31's halo is in clearly identifiable substructure.

Less data exist for the stellar halos of massive early type galaxies. In pioneering work, the {\it Hubble Space Telescope} (\hst) was used to study the stellar halos of the local early type galaxies NGC~3115, NGC~3377, NGC~3379 and NGC~5128 \citep{Elson97, Kundu98, Harris02, Rejkuba05, Harris07a, Harris07b, Rejkuba11} -- initially finding only relatively enriched stars. It was only at much larger radii (11$r_{e}$) that metal-poor stars started to emerge, although this was as a broad metallicity distribution in the halo of NGC~3379, rather than a distinct metal-poor population \citep[][]{Harris07a}. In recent work, more distant fields in NGC~5128 were observed by \hst \citep{Rejkuba14}. This showed a quite enriched halo out to 25$r_{e}$. However, the median halo metallicity did decrease with radius and the outermost field was observed to have a higher fraction of metal-poor stars. 

Another probe of a galaxy's halo is its globular cluster system. These star clusters are associated with major periods of star formation and are therefore thought to provide chemo-dynamical tracers of the underlying stellar populations of a galaxy. This is observed to be the case in the Milky Way and M31, where the metal-poor globular clusters are associated both spatially and chemically with the metal-poor halo population \citep[see, e.g., the review of][]{Brodie06}. Globular clusters are particularly useful as probes of the stellar populations of galaxies because their relatively high surface brightness allows them to be accurately observed in relatively distant galaxies and at large galactocentric radii. These clusters can also contain a relatively large fraction of the stellar halo population of a galaxy -- for example, 2$\%$ of the Milky Way's metal-poor halo mass is in its globular clusters, but these clusters only comprise 0.3$\%$ of the Galaxy's total mass. 

The globular cluster populations of early-type galaxies are observed to have a strongly bimodal metallicity distribution \citep[e.g.][]{Peng06,Strader06}, with the metal-poor globular clusters being more spatially extended than the metal-rich globular clusters \citep[e.g.][]{Bassino06,Brodie06,Forbes11}. These observations suggest the presence of an underlying metal-poor stellar halo that extends to large radii around most early-type galaxies. However, direct detection of a distinct metal-poor stellar halo that corresponds to the metal-poor globular cluster population in early-type galaxies has proved challenging. 

In this paper, we present the stellar populations in the outer halo of NGC~3115 -- a nearby S0 galaxy at a distance of 10.2~Mpc (see Section \ref{sec:TRGB}) with $L_{K} = 9.5\times10^{10} \LKsun$ \citep{Jarrett03} and $r_{e}=2.5$~kpc \citep{Capaccioli87}\footnote{This is the ``bulge $r_{e}$'' that was previously used to investigate the galaxy's globular cluster population \citep{Arnold11, Jennings14}. Along the minor axis of the galaxy $r_{e}$=1.7~kpc \citep{Capaccioli87}. The halo of NGC~3115 may have a lower ellipticity than the inner regions. We therefore choose the more circular ``bulge $r_{e}$'', but note that increased ellipticity places our fields at even larger $r_{e}$ than quoted.}. Importantly, we observe this galaxy's halo at very large galactocentric radii -- where any metal-poor halo should be more dominant. This galaxy is close enough to detect and resolve stars in its halo with the \hst and it hosts a large and strongly bimodal globular cluster population \citep[e.g.][]{Brodie12}. As such, it is a prime location for observing an associated bimodal stellar halo population. Comparison with the galaxy's well studied globular cluster population also provides a strong test of whether these globular clusters truly trace a galaxy's stellar halo.

The data used in this study are discussed in Sections \ref{sec:data} and \ref{sec:photometry}. In Section \ref{sec:TRGB} we use the tip of the red giant branch (TRGB) to determine the distance to NGC~3115. We discuss NGC~3115's stellar halo in Section \ref{sec:halo}, presenting its metallicity distributions (in Section \ref{sec:mdfs}) and radial profile (in Section \ref{sec:radial}). We conclude by comparing our observations with simple chemical enrichment models (Section \ref{sec:models}), the halos of other galaxies (Section \ref{sec:other_gal}), and NGC~3115's globular cluster system (Section \ref{sec:gcs}).

\section{NGC~3115 Observations}
\label{sec:data}

\begin{figure}
 \centering
 \includegraphics[height=86mm,angle=0]{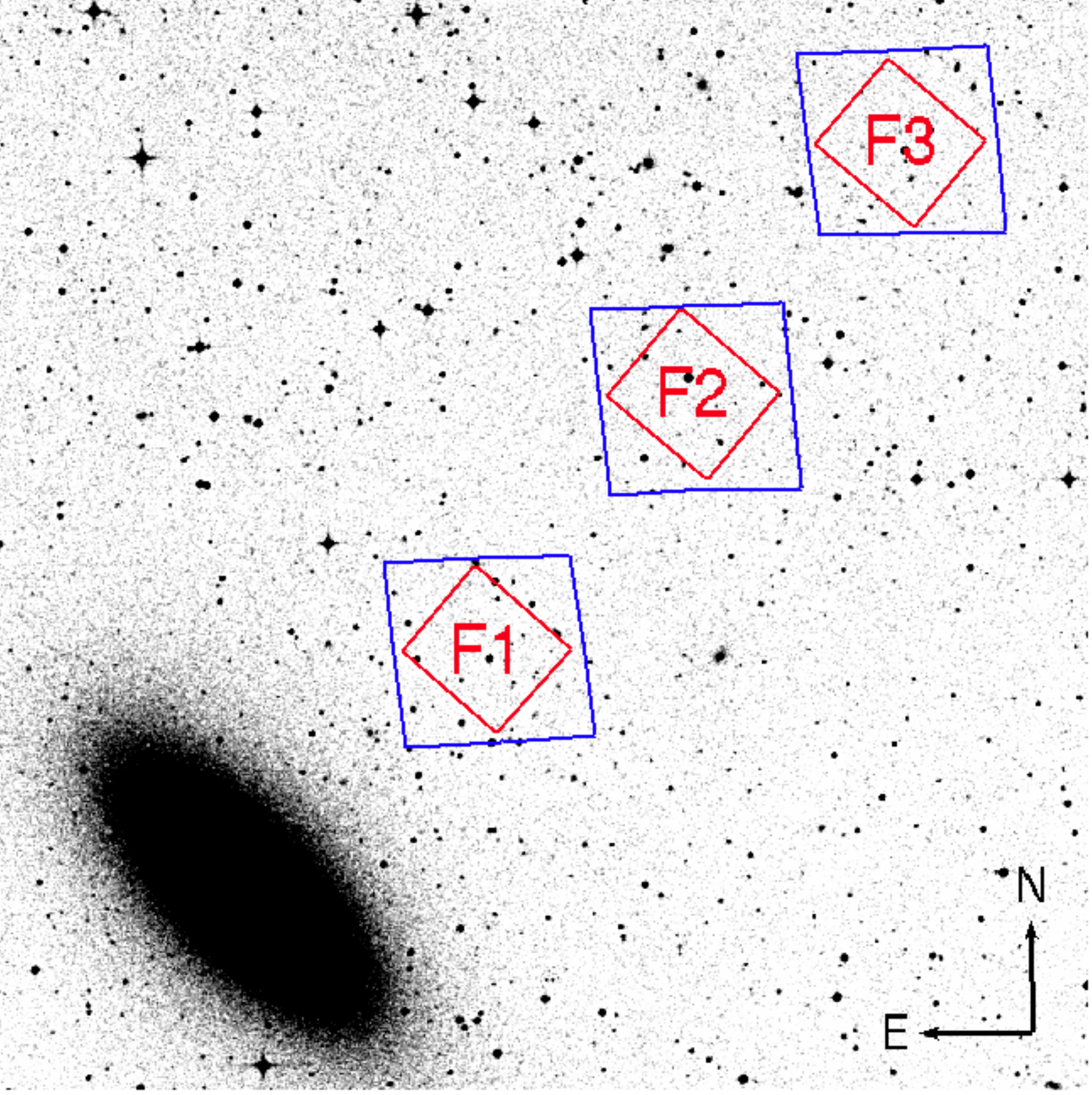} 
 \caption{The fields of NGC~3115's halo that are covered by these ACS/WFC observations (large blue rhombuses) and WFC3/IR observations (smaller red diamonds). The image is taken from the Digitized Sky Survey 2 and covers 20'$\times$20' on the sky. The fields F1, F2 and F3 are at galactocentric distances of 19, 37 and 54~kpc, respectively. This corresponds to 7, 14 and 21~$r_{e}$. \\}
 \label{fig:fields} 
\end{figure}

We obtained \hst observations of three fields in NGC~3115 between October 2012 and May 2013 under the program ID 13048 (P.I. Strader). The fields targeted, shown in Figure \ref{fig:fields}, were chosen to cover the regions of the galaxy where globular cluster observations suggest the stellar population may transition from being relatively metal-rich (in the inner field) to metal-poor in the outer field \citep{Arnold11}. The observations were taken in `coordinated parallel' using the Advanced Camera for Surveys/ Wide Field Camera (ACS/WFC) and the Wide Field Camera 3/ Infrared Channel (WFC3/IR) through the F606W and F110W filters, respectively. The F606W is equivalent to a broad V band filter, while the F110W spans the ground based Y and J bands. 

We utilize the pipeline reduced `flt' images for each WFC3 exposure and the `flc' images for each ACS/WFC exposure, downloading them from the MAST archive\footnote{http://archive.stsci.edu/} in August of 2013. Rather than utilizing the the pipeline drizzle combined images, we combine these flt/flc images together ourselves using the {\sc pyraf/stsdas/drizzlepac} tasks {\sc tweakreg} (to align the images) and {\sc astrodrizzle} (to drizzle combine all of the images together). We run these routines with the default parameters, with the exception of drizzle parameters final$\_$pixfrac=0.8 and final$\_$scale=0.0333 (for the ACS/WFC3) and 0.0666 (for the WFC3/IR). The resulting combined images of fields F1 and F3 have total exposure times in F110W and F606W of 18240s and 18580s, respectively. Field F2 has twice the exposure, with total exposure times in F110W and F606W of 36480s and 37160s, respectively. 

\section{Photometry} 
\label{sec:photometry}

We perform our primary photometry using the ACS and WFC3 specific modules of the {\sc dolphot} software \citep[version 2.0, a modified version of HSTphot;][]{Dolphin00}. Rather than using the combined drizzled images for photometry, {\sc dolphot} is run using individual `flt' images, with the software using the drizzle combined image only as a reference image. The task performs PSF photometry for each exposure and produces a combined magnitude for each source. This includes, the required calibration, aperture corrections, and charge transfer efficiency corrections (for the `flt' images). We use the PSFs that are provided with the {\sc dolphot} software. These are from the TinyTim simulation with Anderson cores. Performing the photometry in this way is computationally more intensive. However, this method can potentially produce more accurate photometry and allows more accurate calculation of errors than using the drizzled images (where the drizzling process produces pixels with correlated noise). 

To select only those sources with reliable photometry, we require the source to have the following {\sc dolphot} flags in both filters: object$\_$type = 1 or 2, which selects stellar sources; error$\_$flag $\leq$3; sharpness in the range -0.3 to 0.3, this helps to remove blended sources, background galaxies and cosmic rays/detector artifacts; and crowding $<$0.3mag, which removes sources whose photometry is significantly affected by neighbors. Stars are detected to 3.0$\sigma$ detection limits of ${\rm F606W}=29.7, 30.4, 30.1$ and ${\rm F110W}=28.4, 28.9, 28.5$ in fields F1, F2 and F3, respectively. The resulting catalogs of sources in the F606W and F110W filters are combined based on their aligned WCS information using the software {\sc stilts/topcat} \citep{Taylor06}. 

We find that {\sc dolphot}'s flags are not sufficient to remove all extended sources from our catalogs. We therefore use {\sc sextractor} to provide a better discrimination for marginally resolved sources. We run {\sc sextractor} over the drizzle combined images and match the resulting catalogs to that produced by {\sc dolphot}. The colors and magnitudes are consistent between our {\sc sextractor} and {\sc dolphot} catalogs. To remove extended sources we require that the {\sc sextractor}'s `class$\_$star' flag be $>$0.5 in the F606W filter. This filter provides a better discriminant than the F110W filter, likely due to the larger pixel size of the WFC3/IR and the shape of the IR PSF. We also remove extended sources by noting that point sources brighten by less than 0.5~mag from a 0.15$\arcsec$ to a 0.3$\arcsec$ aperture for the F606W filter and a 0.2$\arcsec$ to 0.4$\arcsec$ aperture for the F110W filter. Our final reduced (total) catalogs contain 4951 (13914), 2335 (5295) and 549 (2126) sources for fields F1, F2 and F3, respectively. 

\begin{figure}
 \centering
 \includegraphics[height=86mm,angle=270]{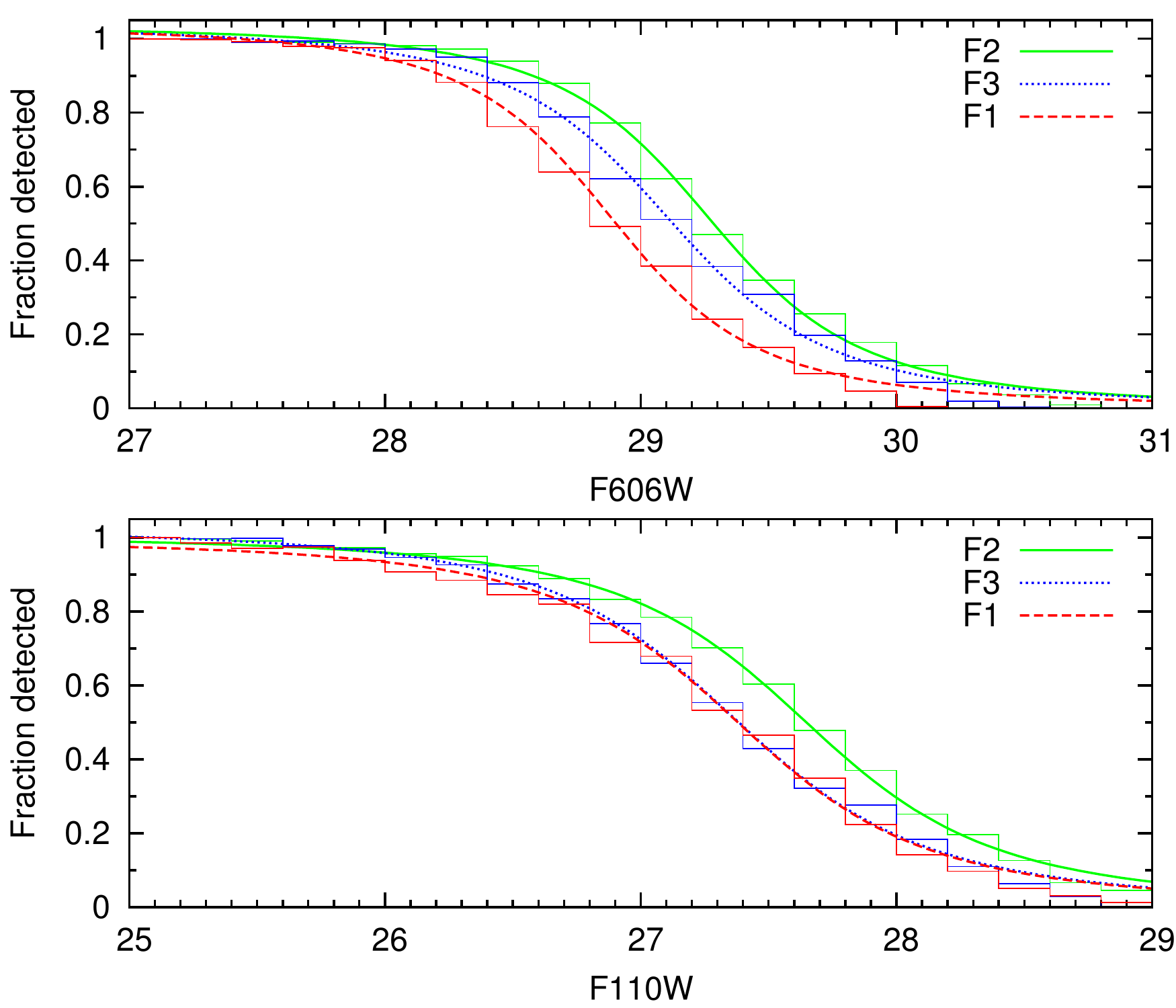} 
 \caption{Fraction of artificial stars detected as a function of F606W (top) and F110W (bottom) magnitude. In both panels we plot the completeness in our three fields: F1 (red-dashed); F2 (green-solid); and F3 (blue-dotted). The curves are scaled to have a completeness of 1 in the brightest magnitude bin. As expected, the longer exposure time of the middle field (F2) results in slightly deeper detection limits. The outer field (F3) also reaches slightly deeper detection limits than the inner field. This is likely due to crowding effects. The curves are from fitting Equation \ref{eq:completeness} to the data. \\}
 \label{fig:completeness} 
\end{figure}

\begin{figure*}
 \centering
 \includegraphics[height=180mm,angle=270]{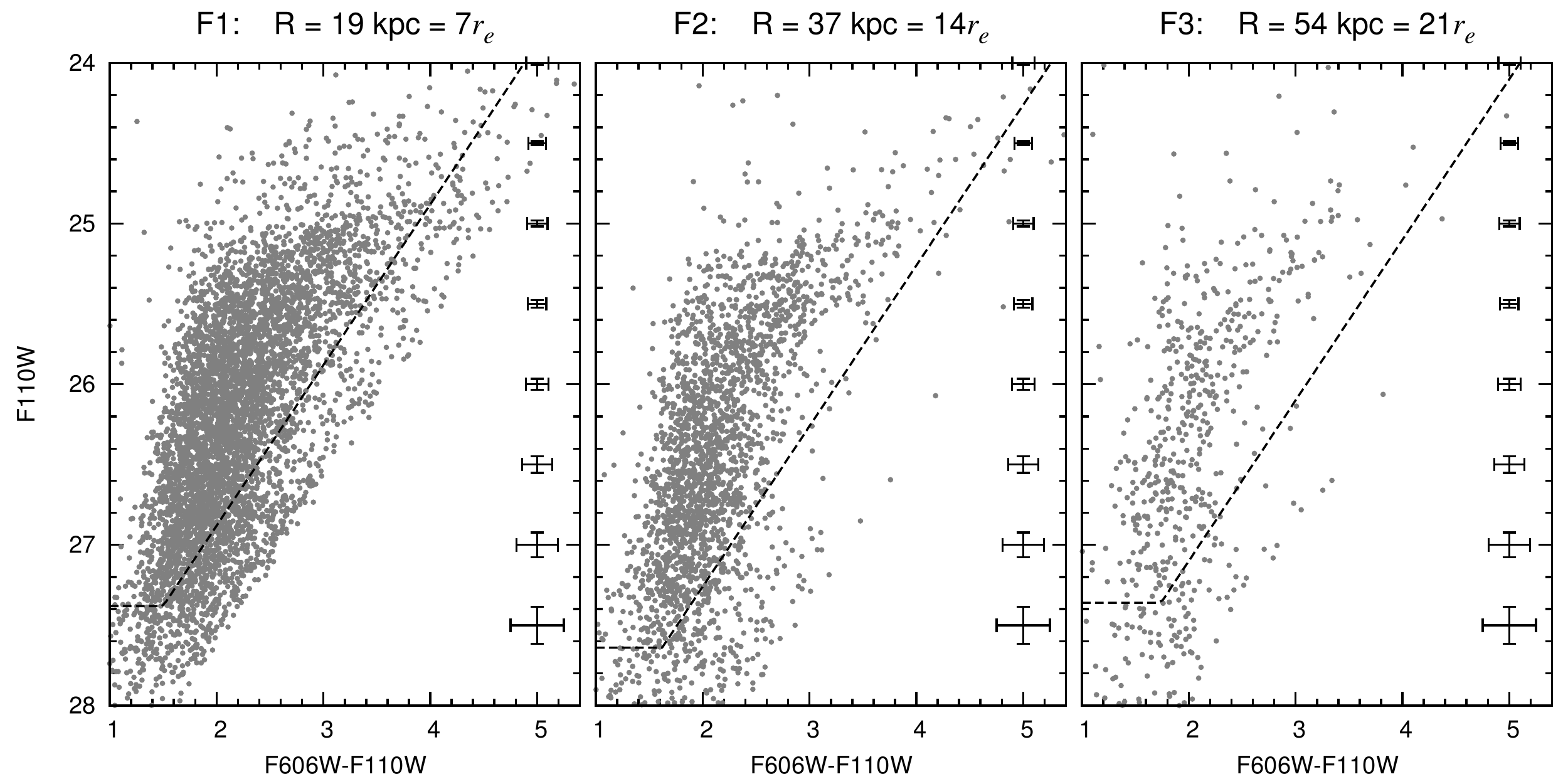} 
 \caption{CMDs for star-like sources from the three fields in the halo of NGC~3115. The fields are located at galactocentric radii of 19~kpc (7$r_{e}$), 37~kpc (14$r_{e}$) and 54~kpc (21$r_{e}$), from left to right, respectively. The dashed line shows the 50$\%$ completeness limit.  \\}
 \label{fig:cmds1} 
\end{figure*}

The detection limits of these data in both filters can be estimated using {\sc dolphot}'s artificial star tests. To do this, we place 40,000 stars in each image and perform the same photometry. For the artificial stars to be classed as detected, we require that the source is detected at $>$3$\sigma$ and that the measured magnitude is within 0.5 magnitudes of that input. Figure \ref{fig:completeness} shows the completeness function for each field through both filters. Field F2 is complete to fainter F606W and F110W magnitudes due to the longer exposure time. For F110W, crowding is not important and the inner and outer fields (F1 and F3) have similar completeness limits. For the F606W observations, crowding results in field F1 having shallower detection limits than the outer field. For the F110W images, only 90-95$\%$ of the artificial stars are detected in the brightest magnitude bin. Stars of this magnitude should easily be detected in these data. The missed sources are generally located close to bright sources and this is the likely reason for their poor/non detection. The incompleteness due to crowding does not have a color dependence that would significantly effect the derived metallicity distribution. Therefore, for our primary analysis of the RGB population, we are not concerned with spatial completeness and can scale the completeness functions so that the completeness in the brightest magnitude bin is 100$\%$. The detection fraction is fit to a completeness function of the form: 

\begin{equation}\label{eq:completeness}
f = \frac{1}{2}\left(1-\frac{\alpha(m-m_{50})}{\sqrt{1+\alpha^2(m-m_{50})^2}}\right)
\end{equation}
\\
where $m_{50}$ is the magnitude at which 50$\%$ of the sources are detected. The resulting 50$\%$ completeness limits are calculated as $m_{50}$(F110W)=27.38, 27.64, 27.36 and $m_{50}$(F606W)=28.88, 29.26, 29.10 for the fields F1, F2 and F3, respectively.

For analysis that relies on the distribution and total number of stars, such as the radial stellar density profile, we calculate two additional completeness corrections. Since the completeness is not constant across the images, primarily due to saturated stars and large background galaxies, we produce a spatial completeness map for each image. This is done based on our artificial star tests. Additionally, we note that our completeness correction is only valid for our {\sc dolphot} photometry. When matching to the {\sc sextractor} catalogs and removing galaxies, it is likely we also remove some stars. This is only found to be significant for the inner field, where crowding results in our star/galaxy separation being less reliable and rejecting some genuine stars. We correct for this by assuming the number of background galaxies is constant across the three images and comparing this to the number of sources removed from the inner field.

\subsection{Color magnitude diagrams of NGC~3115's halo} 

In Figure \ref{fig:cmds1}, we show the CMDs for all star-like sources, with reliable photometry, in the three halo fields of NGC~3115 (F1-3 from left to right). The black-dashed lines in this figure indicate the estimated 50$\%$ completeness limit for these data. The median errors are shown as a function of F110W magnitude. These errors are consistent with those estimated from our artificial star tests. It should be noted that, at fixed F110W magnitude, the uncertainty increases slightly for redder colors (due to the source flux decreasing in the F606W filter). 

As expected, the source density decreases sharply from the inner field (at 7$r_{e}$) to the outer field  (at 21$r_{e}$). However, significant numbers of stars are detected in all three of the halo fields of NGC~3115. A rapid decrease in the density of sources is clearly visible for F110W~$\lesssim 25.1$. This region is associated with the tip of the TRGB. A small number of sources are observed above the TRGB. These can be explained as a combination of several different sources. 

In crowded fields, the blending of sources can result in the detection of a single source brighter than the TRGB. Our halo fields are not very crowded due to the quite low stellar density and the high spatial resolution of the \hst images used. However, it is possible that some objects are still blended together. This effect will increase with stellar density -- and hence the square of the number of stars. We estimate the influence of blending on our photometry using our artificial star tests. This confirms that blending is insignificant in the two outer fields (F2 and F3). In the inner field (F1), where the stellar density is higher, a tail of stars extending to 0.4 magnitudes above the TRGB is observed in our artificial star tests. This is only a small fraction of the total number of stars in the field, but these blends may produce over half of the sources observed above the TRGB. 

Some stars in the NGC~3115's halo can have luminosities in excess of the TRGB, such as long period variable (LPV) stars. These LPV stars are likely to be dominated by TP-AGB stars, but some variable RGB stars may also reach these magnitudes. Such bright stars are more prevalent in intermediate age populations and will have relatively short lifetimes \citep[e.g.][]{Dalcanton12b}. However, some of these stars should be present in an old stellar population \citep{Renzini98}, as is observed in old metal rich globular clusters \citep[e.g.][]{Guarnieri97, Lebzelter14} and in other nearby galaxies \citep[e.g.][]{Harris07b, Girardi10}. The number of these bright stars will scale with the size of the stellar population and hence the number of RGB stars. 

There is also likely to be some contamination from foreground stars and (unresolved) background galaxies in the direction of our fields. Because our fields cover a similar area and direction on the sky, this background should be similar between the three fields. Using the Besan\c{c}on model of the distribution of stars in the Milky Way \citep{Robin03}, we expect only a few foreground stars in the magnitude and color range above the TRGB. Similarly, at these magnitudes, we would expect only a few unresolved background galaxies to be present \citep{Radburn-Smith11}. 

To estimate the relative contributions of these different bright sources to our three fields, we fit the number of stars observed above and below the TRGB to a function composed of: a constant, representing contamination from foreground stars and background galaxies; a linear component, representing the LPV stars that scale with the number of RGB stars; and a component due to blending, which varies with the square of the number of RGB stars and is scaled to fit to the artificial star observations. We conclude that foreground stars/ background galaxies may account for $\sim 10$ bright sources per field (consistent with the small numbers expected). The fraction of bright LPV stars to RGB stars (within 0.7 magnitudes of the TRGB) is estimated at $\sim2.0\%$ \citep[consistent with the fractions observed in other galaxies by][]{Girardi10}. The bright sources in the two outer fields (F2 and F3) are dominated by these bright stars and background contamination. The inner field (F1) contains more stars and we find that blends may account for over half the bright stars observed, with LPV stars being the other significant component. We note that the number of bright stars in all fields is a small fraction of the number of RGB stars observed.

\section{The distance to NGC~3115} 
\label{sec:TRGB}

\begin{figure}
 \centering
 \includegraphics[height=86mm,angle=270]{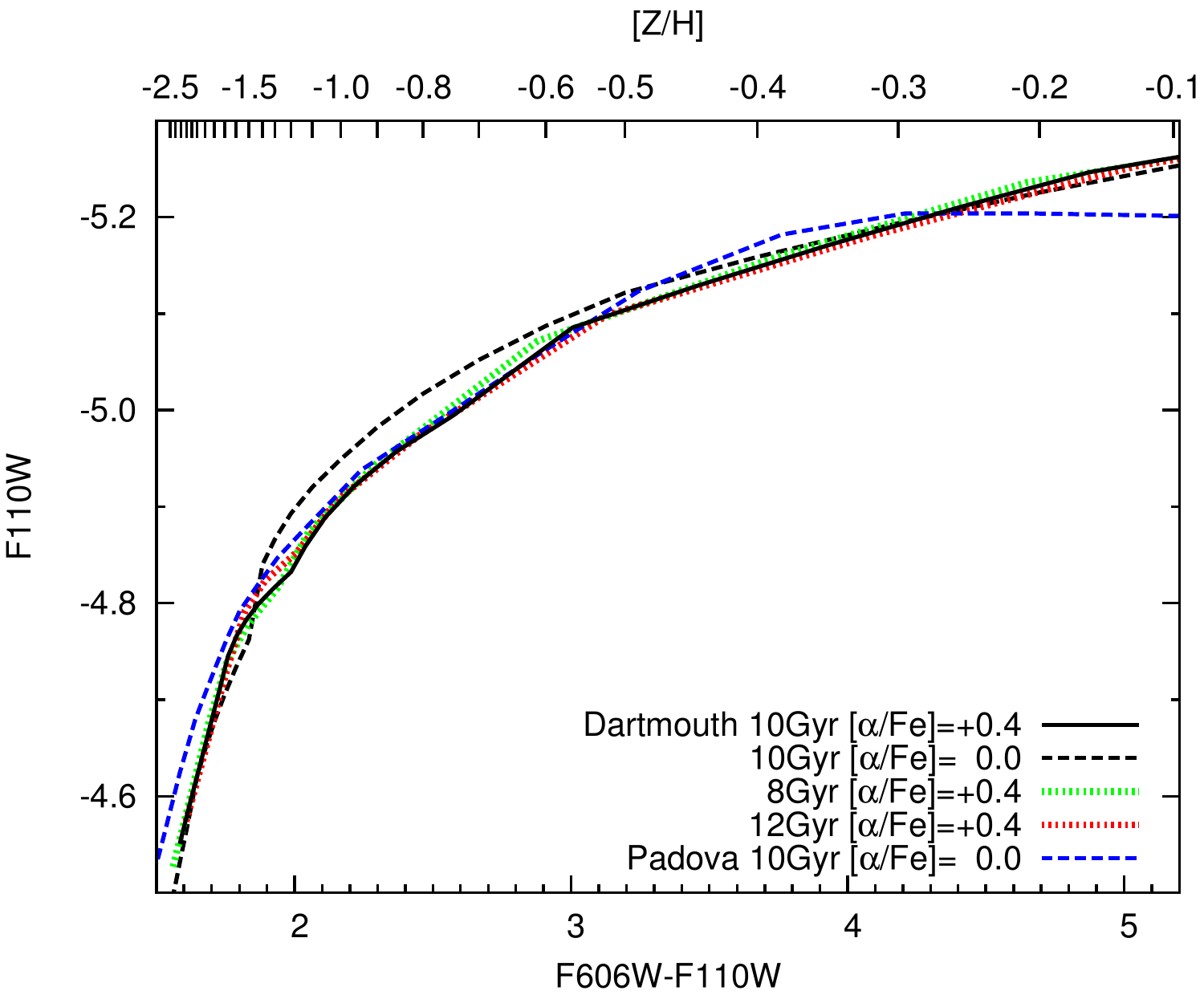} 
 \caption{The absolute F110W magnitude of the TRGB as a function of color (and corresponding \Z on the top axis). The solid-black line is from the Dartmouth isochrones for a 10~Gyr old stellar population with ${\rm [\alpha/Fe]} = 0.4$. This is the model used to detrend the data in our TRGB analysis. We also plot curves for the same model but with ${\rm [\alpha/Fe]} = 0.0$ (black-dashed) and ages of 8~Gyr (green-dotted) and 12~Gyr (red-dotted). The dashed blue line shows a 10~Gyr stellar population from the Padova isochrones.  It can be seen that no significant offsets are found between the different models or stellar populations and that: the age has little influence over the range 8-12~Gyr; the $\alpha$-element abundance has a small effect ($<$0.05) that is largest at intermediate \Z; the Dartmouth values are in good agreement with the Padova values, with the largest difference ($\sim$0.05) at the extremes of the metallicity distribution. \\}
 \label{fig:TRGB_mag} 
\end{figure}

\begin{figure}
 \centering
 \includegraphics[height=86mm,angle=180]{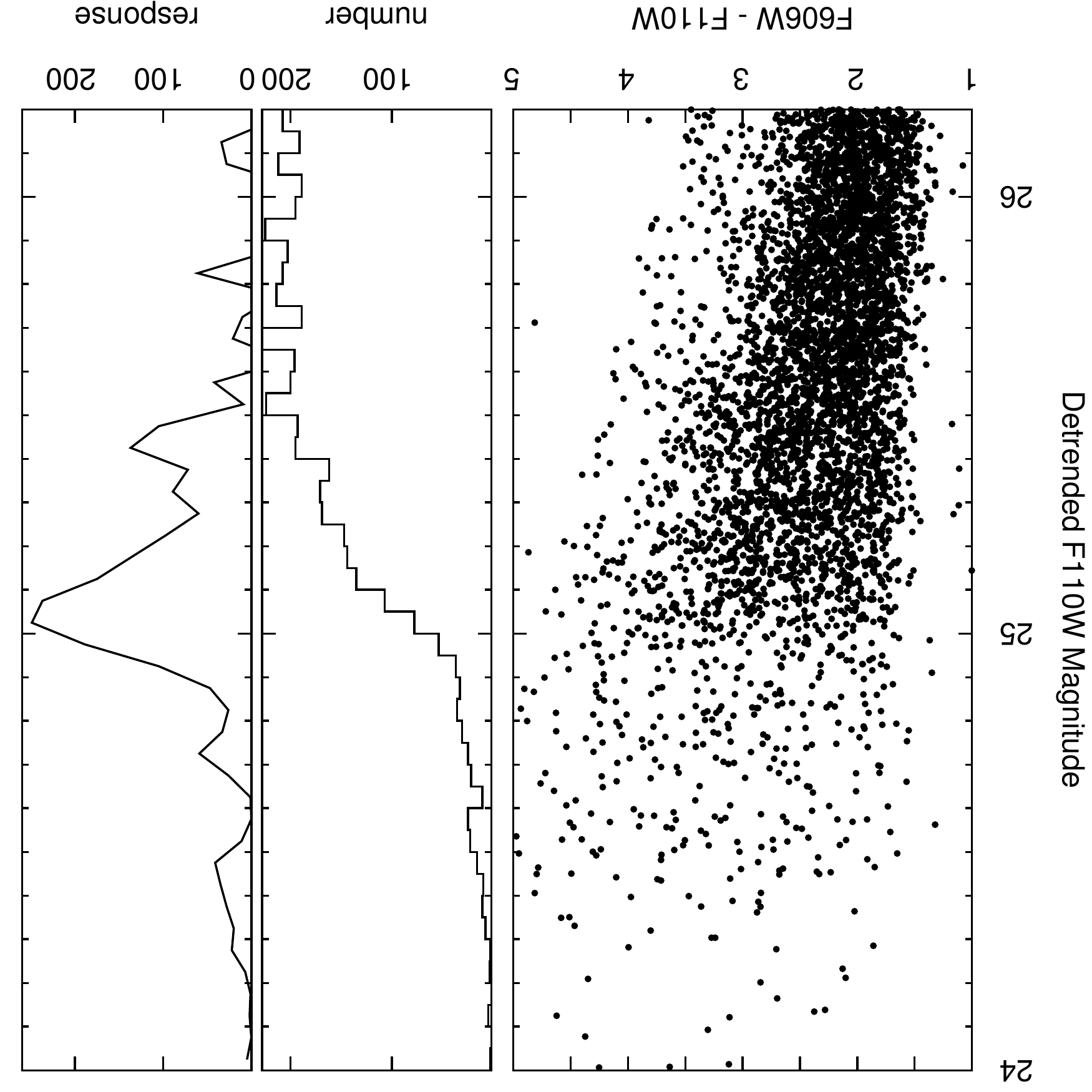} 
 \caption{{\it Left:} The combined CMD of all three fields studied. The F110W magnitude has been detrended so that the TRGB is constant as a function of color. This detrending appears to accurately represent the observed population, with a rapid decline in the density of sources with F110W~$\lesssim$~25 clearly apparent. {\it Middle:} A histogram of the number of sources at each magnitude (with 0.05~mag bins). {\it Right:} The response from the Sobel filter -- values of zero suggest a flat source density while large values suggest a rapid increase in the number of sources. A peak is clearly apparent at F110W~=~25.05. \\}
 \label{fig:TRGB_ngc3115} 
\end{figure}

Clearly visible in Figure \ref{fig:cmds1} is a rapid decrease in the density of sources with F110W~$\lesssim$~25.1. This is associated with the TRGB and can be used to determine an accurate distance to NGC~3115. The only previous TRGB distance to NGC~3115 was obtained from a single WFPC2 field. This was analyzed by two independent studies which found the distance modulus to be 30.21$\pm$0.25 \citep{Elson97} and $30.19\pm0.3$ \citep{Kundu98}. The data presented in this study cover more area, are less crowded, and are much deeper than those of that previous study -- this allows us to more precisely constrain the TRGB location. 

\subsection{The absolute F110W mag of the TRGB}

A complication with using our data to measure a TRGB distance to NGC~3115 is that, in both the F606W and F110W filters, the magnitude of the TRGB varies as a function of metallicity -- as opposed to previous work in the I-band, where variations in the TRGB, as a function of metallicity, are relatively small \citep[e.g.][]{Lee93}. We take the TRGB magnitude in the F110W filter from the Dartmouth Stellar Evolution Database isochrones \citep{Dotter08} for a 10~Gyr stellar population with [$\alpha$/Fe]=0.4. We plot the TRGB as a function of \Z in Figure \ref{fig:TRGB_mag}. It can be seen that the TRGB increases by around 0.6~mag from $\Z = -2.2$ to $\Z = -0.1$. However, the predicted variation is similar for different models and model parameters. Only small variations in the TRGB magnitude are produced by altering the age of these models by $\pm$2~Gyr, or the [$\alpha$/Fe] by 0.4. Additionally, good agreement is found between these Dartmouth isochrones and the Padova isochrones of \citet[][with age~=~10~Gyr and ${\rm [\alpha/Fe]}=0.0$]{Marigo08}. The absolute magnitude of the TRGB in F110W is quite poorly constrained empirically and thus we rely on these model isochrones. For now, we estimate a systematic error on the TRGB magnitude of 0.1~mag \citep[consistent with the uncertainties found by the NIR TRGB study of][]{Wu14}.

We choose the 10~Gyr Dartmouth isochrones with  [$\alpha$/Fe]~=~0.4 to account for the variation in the TRGB as a function of F606W--F110W color. These parameters are consistent with those expected in NGC~3115's halo \citep[see, e.g., the integrated spectroscopy out to 5~kpc along NGC~3115's minor axis,][]{Norris06}. This model (and all those plotted in Figure \ref{fig:TRGB_mag}) have been reddened to simulate the Galactic extinction in the direction of NGC~3115, where ${\rm E(B-V)}=0.042 \pm 0.001$ \citep{Schlafly11}. We use this model to correct the stellar photometry so that the TRGB is at F110W~$=-5$ for all colors. The resulting CMD is plotted in the left panel of Figure \ref{fig:TRGB_ngc3115}. Here, we combine the photometry from all three fields. We do not expect a significant change in the distance of the stars in the stellar halo and note that line-of-sight depth effects are insignificant since a depth of 20~kpc at this distance results in a magnitude variation of only 0.005. Therefore combining these fields helps to increase the accuracy of our TRGB determination by increasing the sample size. Comparing this CMD to those in Figure \ref{fig:cmds1}, it can be seen visually that the TRGB is much more constant as a function of color -- this verifies that the model isochrones used to detrend the data provide a good representation of the variation of the TRGB with color.

\subsection{The TRGB in NGC~3115}

To determine the TRGB location in this CMD, we run a Sobel filter over the luminosity function plotted in the middle panel of Figure \ref{fig:TRGB_ngc3115}. This filter is known to be a robust edge detection algorithm \citep[see e.g.][]{Lee93}. We use an expanded version of the Sobel filter, which works by moving a filter of the form $[-p_{-2},-p_{-1},+p_{+1},+p_{+2}]$ across the data. Here, $p_{-2}, p_{-1}, p_{+1}$ and $p_{+2}$ are the values one and two before and after the current value. In this way, flat distributions will cancel out and have a value of zero, while steep positive gradients produce a strong signal. The right panel of Figure \ref{fig:TRGB_ngc3115} shows the response from this filtering as a function of magnitude. The strongest response is at F110W~=~25.025. This is clearly associated with a sharp drop in the source density in the left panel and is taken to be the TRGB. 

Our method for finding the TRGB relies on binning the data; a bin width of 0.05 mag is used. We experimented with using smaller bin sizes to increase our magnitude resolution, but the increased noise resulted in a weaker responses from the Sobel filter. To investigate the accuracy of the TRGB magnitude obtained, we used a bootstrapping method. This method produces a new dataset from our original data by randomly selecting entries from it (with replacement and equal probability) until the new dataset is the same size as the original. We then rerun our analysis on the new data and measure the resulting TRGB. We repeat this process $10^{5}$ times. The standard deviation of the resulting TRGB estimates from our bootstrapped runs is 0.05 and the mean is slightly shifted, with the TRGB at F110W=25.048. This shift is likely due to the binning of the data, which limits our real data run to a 0.05~mag grid. 

From this analysis, we take the TRGB to be at F110W~$=25.05\pm0.05$. The data have been calibrated to define the ${\rm M_{TRGB}}=-5.0$, with systematic errors due to the model used and stellar population effects estimated as $\pm$0.1. We therefore obtain a distance modulus for NGC~3115 of $m-M=30.05\pm0.05\pm0.1_{sys}$ (or a distance of $10.2\pm0.2\pm0.5_{sys}$~Mpc).

\subsection{Comparison with other distance estimates} 

Only two previous TRGB distances to NGC~3115 are available in the literature. These are from the same single WFPC2 observation, and yielded consistent distance moduli of $(m-M)=30.21\pm0.3$ \citep{Elson97} and $30.19\pm0.3$ \citep{Kundu98}. Our measurement suggests a slightly closer distance, but is consistent with these previous estimates. We note that our more precise measurement is due to the larger number of stars and lower crowding in our new observations. 

Our derived distance is also consistent with other methods for deriving the distance to NGC~3115: planetary nebulae luminosity function (PNLF) $m-M=30.02\pm0.15$ \citep{Ciardullo02}; surface brightness fluctuations (SBF) $m-M=29.93\pm0.09$ \citep{Tonry01}; and globular cluster luminosity function (GCLF) $m-M=30.00\pm0.07$ \citep{Kundu01}.

\section{The stellar halo of NGC~3115} 
\label{sec:halo}

\subsection{CMDs of NGC~3115's stellar halo} 
\label{sec:cmds} 

\begin{figure*}
 \centering
 \includegraphics[height=180mm,angle=270]{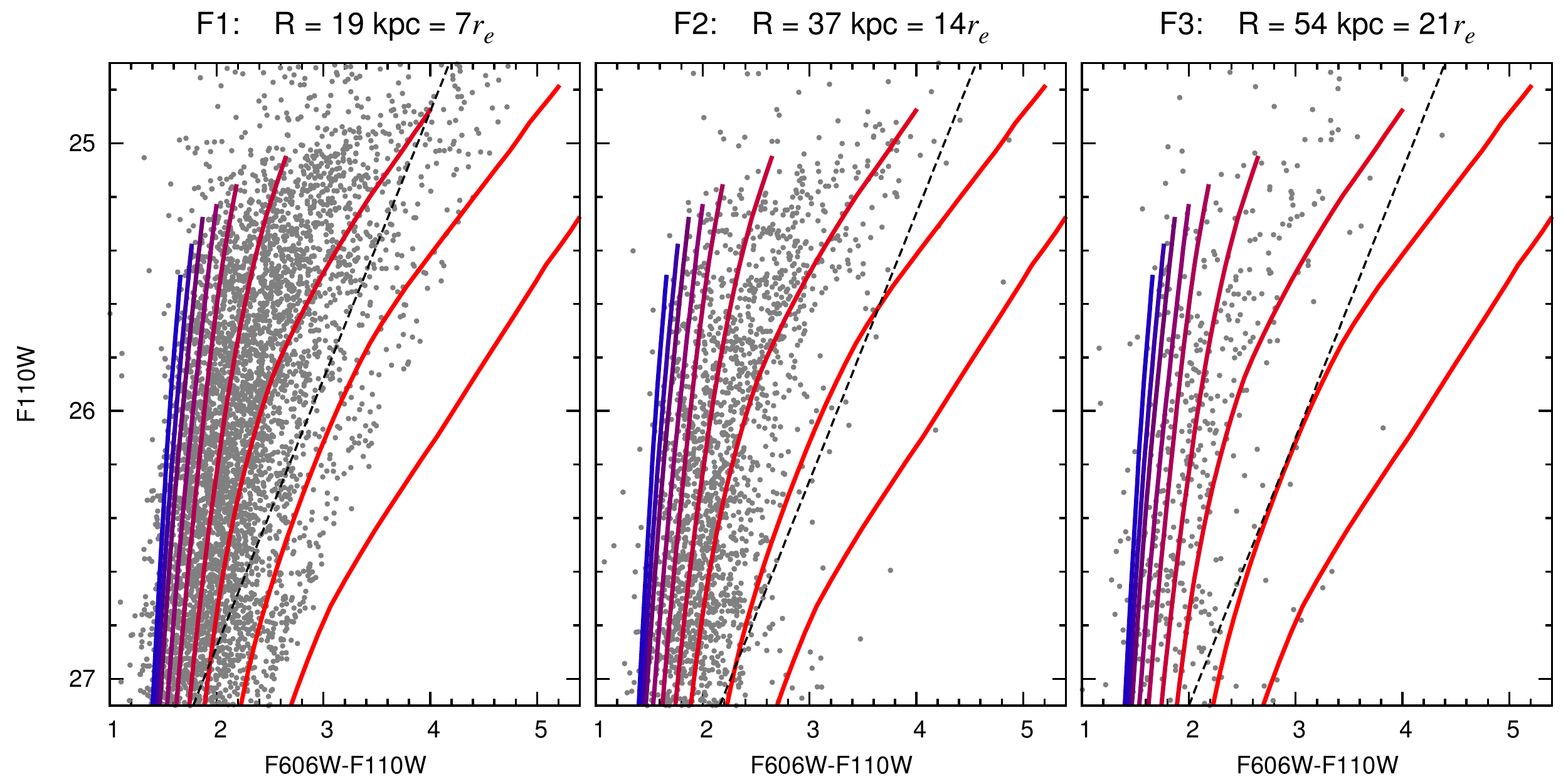} 
 \caption{CMDs for our three fields. The dashed black lines indicate the 50$\%$ completeness limit. The solid lines are theoretical isochrones from the Dartmouth models. These show the tracks of a 10~Gyr stellar population with [$\alpha$/Fe]=+0.4 and metallicities (from left to right and blue to red) of $\Z = -2.2,-1.9,-1.6,-1.3,-1.0,-0.7,-0.4,-0.1,+0.2$. The halo population is clearly in good agreement with these isochrones, but spans the full metallicity range. The sources above the TRGB are likely AGB stars in the halo.}
 \label{fig:cmds2} 
\end{figure*}

\begin{figure*}
 \centering
 \includegraphics[height=180mm,angle=270]{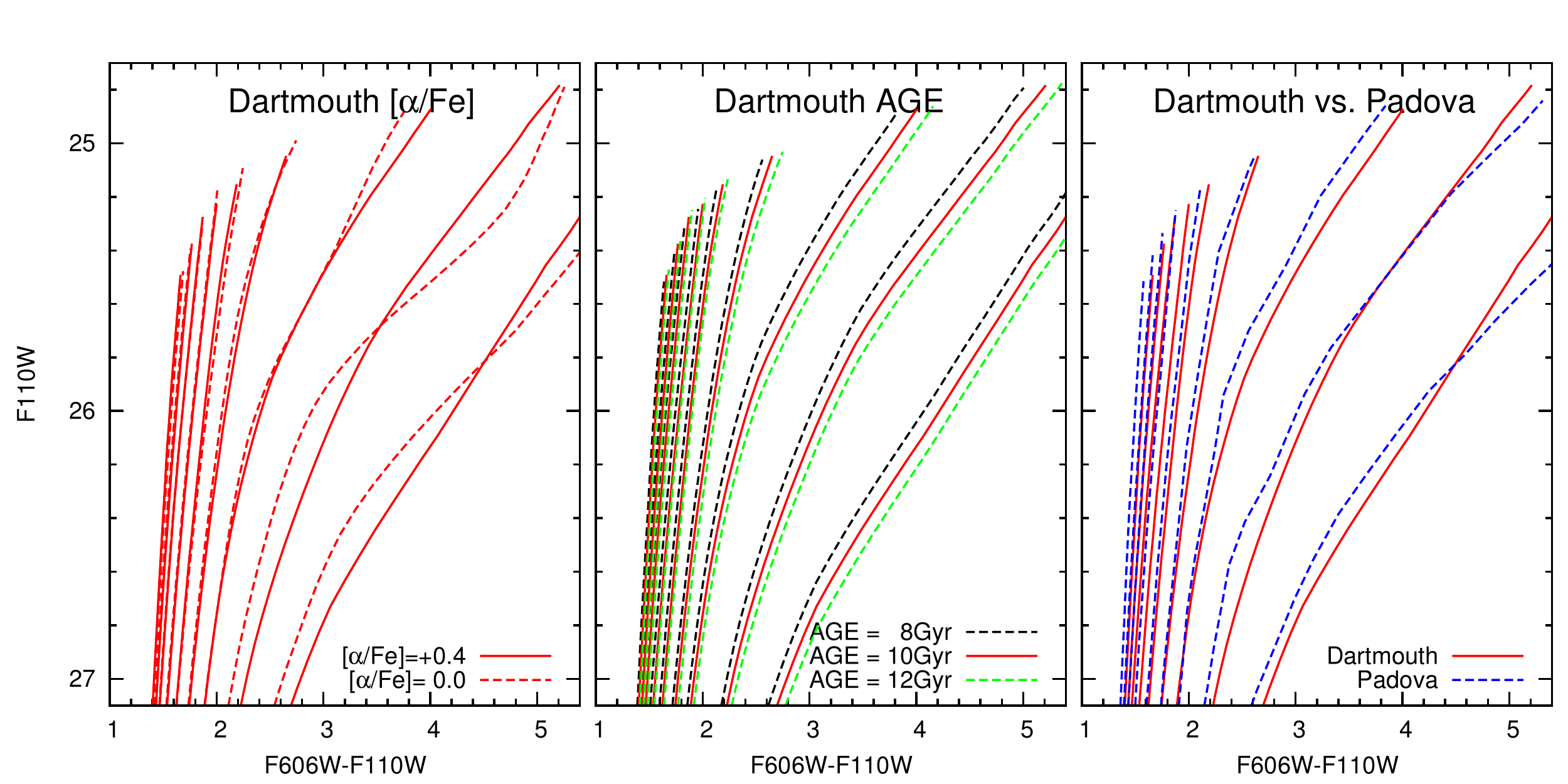} 
 \caption{A comparison of isochrones with slightly different parameters. In all panels the solid red lines are the same isochrones as plotted in Figure \ref{fig:cmds2}. The dashed lines show isochrones with the same \Z, but with a varied second parameter. In the left panel we show the effect of changing the $\alpha$ abundance from +0.4 to 0.0 (solar). It can be seen that this has only a small effect. In the middle panel, we vary the age of the isochrones, showing 8 (black), 10 (red) and 12~Gyrs (green). Again, at these old ages, such variations have only a small effect. The right panel compares the isochrones from the Dartmouth models and the Padova models. Good agreement is found between the different models. The differences observed are generally comparable to (or smaller than) the errors on our photometry. }
 \label{fig:cmd_models} 
\end{figure*}

In Figure \ref{fig:cmds2}, we again show the CMDs for all star-like sources, with reliable photometry, in the three halo fields of NGC~3115. The black dashed lines in this figure indicate the estimated 50$\%$ completeness limit. 

Overlaid on these CMDs are the Dartmouth stellar isochrones for a distance modulus of 30.05 (see Section \ref{sec:TRGB}), an age of 10~Gyr and $[\alpha/{\rm Fe}]$ of +0.4. These parameters are consistent with those expected for a stellar halo population and they are similar to those deduced from integrated spectroscopy of the galaxy's inner halo \citep{Norris06}. We redden the isochrones to account for the Galactic reddening in the direction of NGC~3115, E(B-V)~$=0.042 \pm 0.001$ \citep{Schlafly11}. This reddening corresponds to extinctions of A(F606W)=0.120 and A(F110W)=0.042 \citep{Cardelli89}. From left to right, these isochrones show \Z from -2.2 to +0.2 in increments of 0.3~dex. We convert [Fe/H] to \Z for these $\alpha$-element enhanced isochrones using equation 3 of \citet{Salaris93}. 

These isochrones, evenly spaced in metallicity, show the clear non-linearity between color and metallicity. The metallicity has a much greater effect on the stellar colors at higher metallicities and luminosities. This causes the brighter metal-rich stars to fall beneath our detection limit. At low metallicity, the change in color is much smaller, and the stars bunch to similar colors. The result of this is that the CMDs can appear misleading, with the density of sources increasing artificially towards bluer colors.  

The colors are clearly in excellent agreement with these isochrones for RGB stars in NGC~3115. We note that, while the TRGB is clearly visible as a dramatic decrease in the number of sources, some sources with brighter magnitudes are detected. Many of these are likely to be AGB stars. We do not include the sources brighter than the TRGB in our subsequent analysis. 

In Figure \ref{fig:cmd_models}, we consider the effect of inaccuracies in the choice of isochrones used to represent the stellar population. For $\Z<-0.2$, varying [$\alpha$/Fe] from 0.0 (solar) to +0.4 has a negligible effect on the isochrones for constant \Z. Similarly, varying the age of the stellar population from 8~Gyr to 12~Gyr has only a small effect on the isochrones. Finally, we consider differences between the Dartmouth (red, solid) and Padova (blue, dashed) isochrones. It can be seen that the two models produce very similar colors, with only small differences. We conclude that our analysis is relatively robust to the choice of model and our assumptions about the age and $\alpha$ abundance. Such effects are likely to produce systematic errors in the resulting metallicities of $\sim$0.1~dex. Since much of our analysis is comparative between the three radial fields, systematic errors will have little effect on our conclusions.

\subsection{Metallicity distribution of NGC~3115's stellar halo} 
\label{sec:mdfs} 

\begin{figure}
 \centering
 \includegraphics[height=86mm,angle=270]{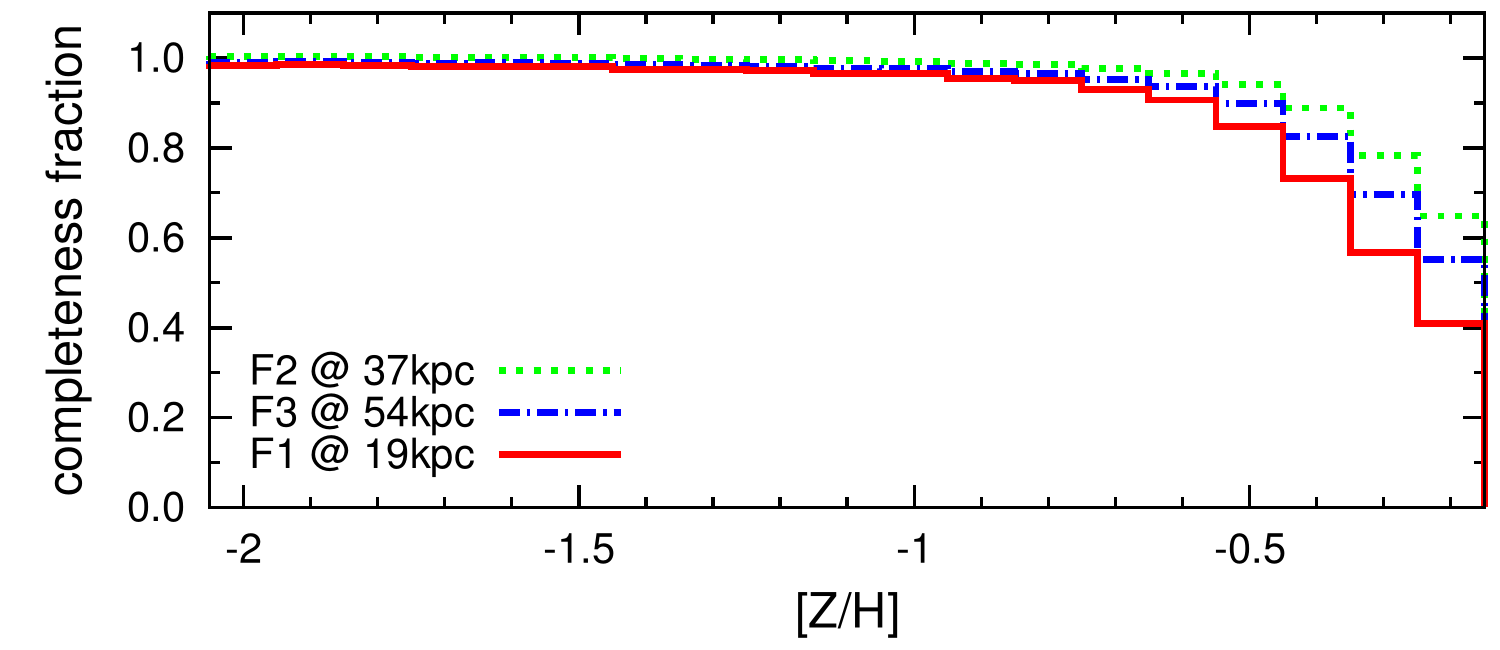} 
 \caption{The completeness as a function of metallicity for stars with F110W~$< 26.5$. The red solid line is for the inner field (F1), the green dotted line is for the middle field (F2) and the blue dot-dashed line is for the outer field (F3). \\}
 \label{fig:MDF_completeness} 
\end{figure}

\begin{figure}
 \centering
 \vspace{-2mm}
 \includegraphics[height=86mm,angle=270]{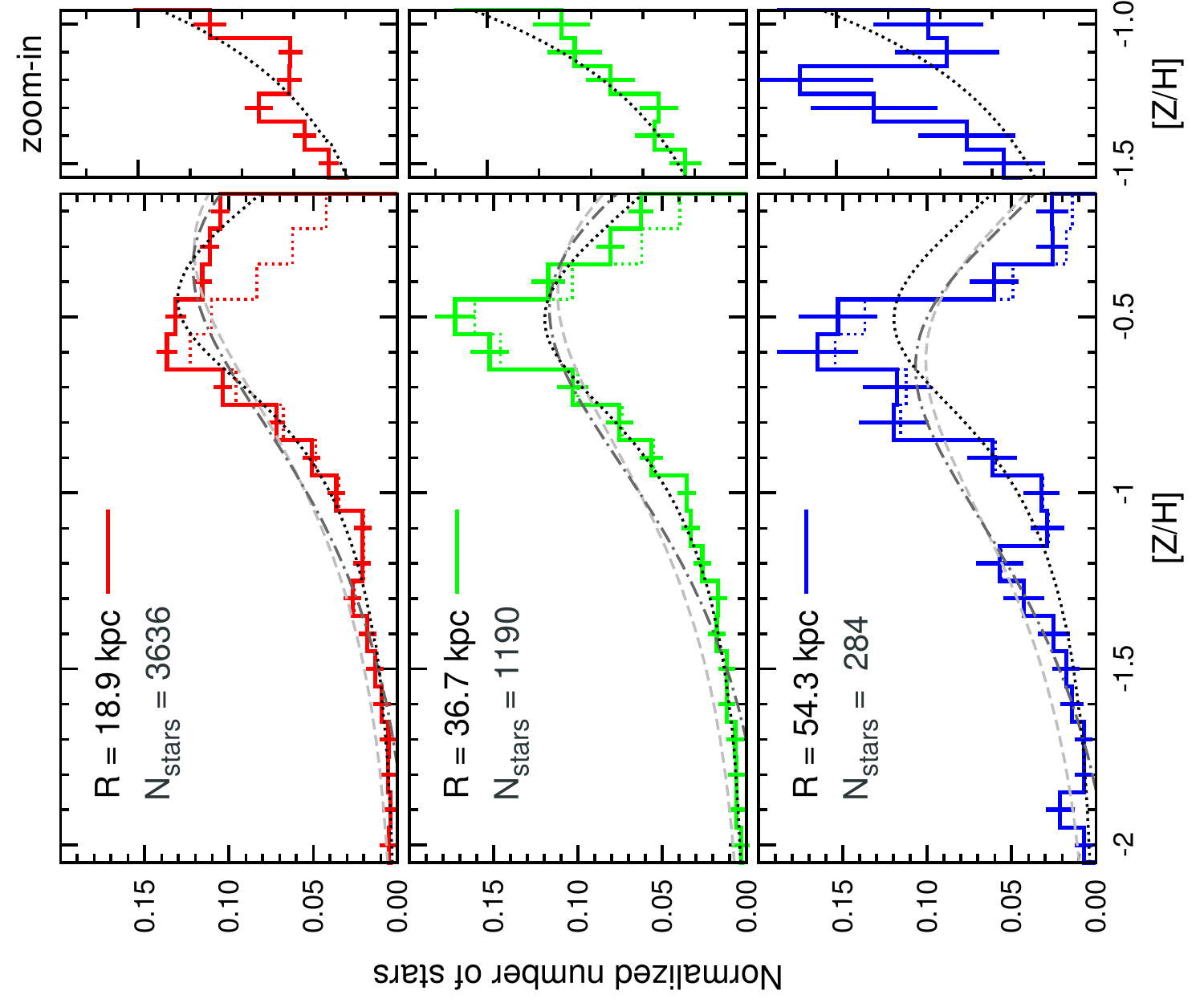} 
 \caption{MDFs for the inner field (F1, top), middle field (F2, middle) and outer field (F3, bottom). The dotted histograms show the observed MDFs before correcting for completeness. The number of stars in each field is normalized to 1. The right panels show zoomed in sections around the metal-poor peak. The errorbars are the associated Poisson errors for each bin. In all fields a relatively metal-rich population with $\Z \sim -0.5$ is present. However, the peak of this population is found to shift to lower metallicities with increasing distance from the center of the galaxy. A second lower metallicity peak can be seen in the tail of the distribution of fields F1 and F3 at $\Z \sim -1.3$. The dashed lightgray, dot-dashed darkgray and dotted black lines show model MDFs from a closed-box, a pre-enriched closed box and an accreting gas model, respectively (see Section \ref{sec:models}). \\}
 \label{fig:MDF_single} 
\end{figure}

\begin{figure}
 \centering
 \vspace{0mm}
 \includegraphics[height=86mm,angle=270]{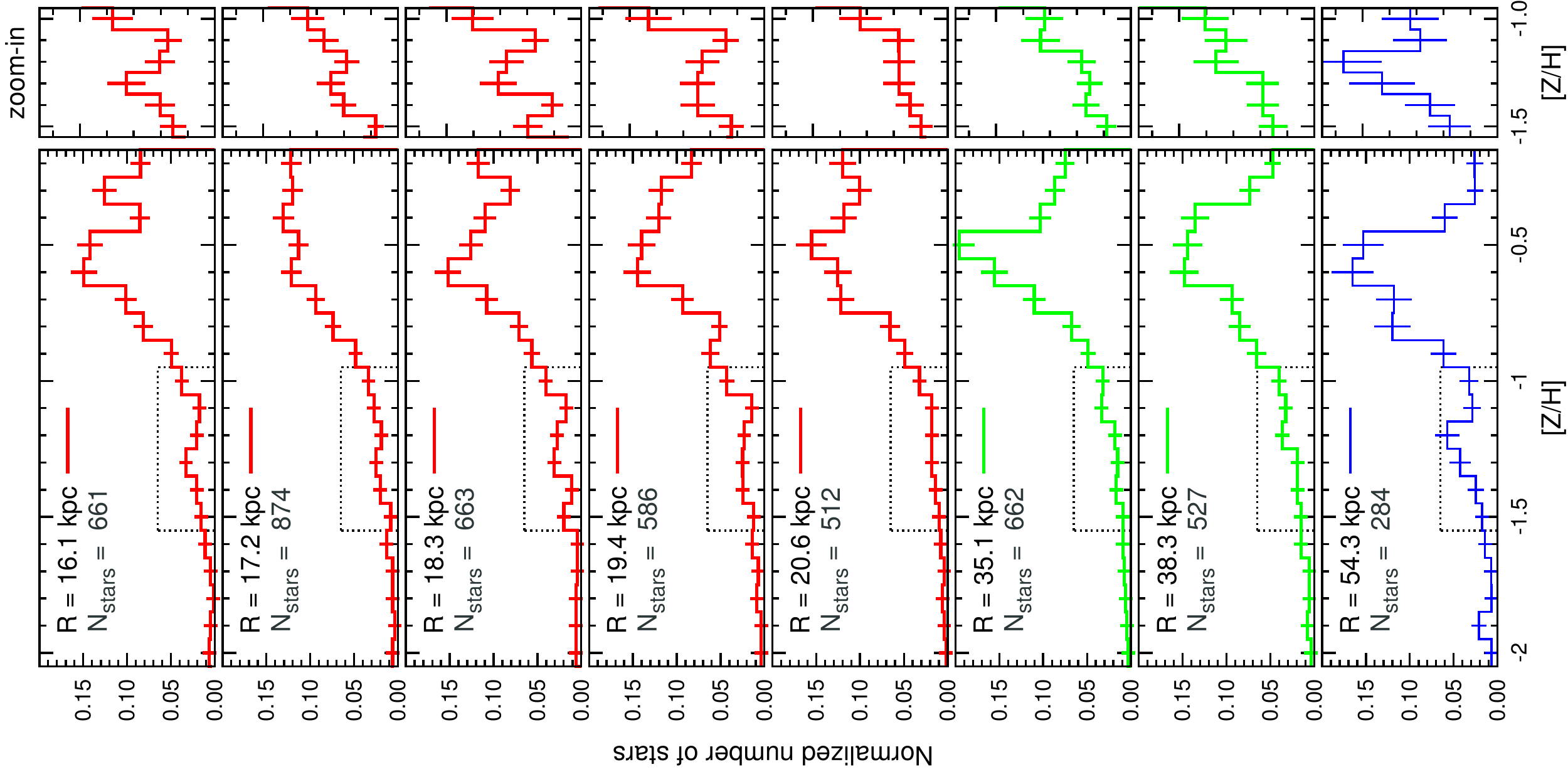} 
 \caption{Same as Figure \ref{fig:MDF_single}, but with the two inner fields split in to radial bins. Distance from the center of the galaxy increases from top to bottom. Despite the increased noise, the low metallicity peak is still observed in the four inner bins. In the middle field the low metallicity peak is not observed at $\sim$35~kpc and the metal-rich population appears more peaked than in the other regions. Further out in this field at $\sim$39~kpc the metal-rich population broadens again and there are hints of the lower metallicity peak. This is suggestive of substructure in the halo at R$\sim$35~kpc. Despite having only half the stars of the other regions plotted, the outer field still shows the strongest low metallicity population -- a likely result of the ratio of metal-poor to metal-rich stars increasing with radius. }
 \label{fig:MDF_rbins} 
\end{figure}

The CMDs presented in Figures \ref{fig:cmds1} and \ref{fig:cmds2} can be used to investigate the stellar metallicity distribution function (MDF) in these three halo fields. To do this, we assume that the stellar population is well represented by the Dartmouth isochrones with an age of 10~Gyr and [$\alpha$/Fe]=+0.4. We then match a star's CMD location to that of a grid of these isochrones separated in \Z by 0.1~dex. As discussed above, small deviations in the age, $\alpha$-abundance and model used should only introduce small differences in the derived metallicities. 

To improve the statistics, the goal is to include as many stars as possible in the derivation of the MDF. However, the increased errors and narrowing of the color difference with metallicity for fainter stars act to artificially smooth out the underlying MDF. The choice of magnitude limit is therefore non-trivial. In the Appendix, we investigate the MDFs produced using  different magnitude cuts. While small variations are observed between the MDFs derived (see Figure \ref{fig:MDF_multi}), the key features are quite insensitive to these choices. In the subsequent analysis we consider all stars with F110W~$<$~26.5. This simple cut allows us to consider a large number of stars with sufficiently accurate photometry. 

Before studying the MDFs of NGC~3115's halo, we first calculate completeness corrections based on the artificial star tests discussed in Section \ref{sec:data}. For stars with F110W~$<$~26.5, this correction is shown in Figure \ref{fig:MDF_completeness}. It can be seen that the completeness correction is only significant for $\Z > -0.4$ and is largest for the inner field (where crowding lowers the detection of these relatively metal-rich stars). We correct the observed MDFs using these completeness corrections and restrict our analysis to stars with $\Z < -0.15$ (because at higher metallicities the completeness corrections become increasingly large and unreliable). 

Figure \ref{fig:MDF_single} shows the completeness corrected MDFs for the three fields observed in NGC~3115's halo (centered at 19, 37 and 54~kpc from the center of the galaxy). The black and gray lines in this figure show the chemical enrichment models discussed in Section \ref{sec:models}. It can be seen that all three fields are dominated by a quite enriched stellar population. The peak in the MDF shifts from $\Z\sim-0.5$ in the inner field to $\Z\sim-0.6$ in the outer field. We also note that the fraction of stars with high metallicity ($\Z >-0.4$) is much lower in the outer two fields than in the inner field. We confirm this variation at the metal-rich end of the MDFs by running $\chi^{2}$ tests between the three different fields over the range $-0.95 < \Z < -0.15$. The resulting $\chi^{2}/\nu$ statistics between the MDFs of fields F1--F2, F1--F3 and F2--F3 are 6.7, 16.6 and 7.8, respectively. For the 8 degrees of freedom, this confirms that the MDFs at all three radii are significantly different at $>6\sigma$. 

All three fields show a tail of stars extending to the lowest metallicities. We define a boundary between the `metal-rich' and `metal-poor' stars to be at $\Z=-0.95$, based on the edge of the main metal-rich peak. The fraction of metal-poor stars is similar in fields F1 to F2 (centered at 19 and 37~kpc). For these fields, the metal-poor stars comprise only 17$\pm1\%$ and 18$\pm1\%$ of the total population, respectively. However, the fraction of metal-poor stars in the outer field (F3, at 54~kpc) is significantly higher at $28\pm3\%$ of the total population. 

A significant feature in the low metallicity tails of the MDFs of fields F1 and F3 is a second peak at $\Z=-1.3$ and $\Z=-1.25$, respectively. The panels on the right of Figure \ref{fig:MDF_single} show a zoomed in view of this section of the MDF. To estimate the significance of this peak, we compare it to the best fitting `accreting gas' model (dotted-black line, see Section \ref{sec:models}). Over the region $-1.45<\Z<-1.15$, we find that this peak is 4$\sigma$ above the model prediction for both fields F1 and F3. This peak is most pronounced in the outermost field, whose MDF has a significantly higher fraction of stars over the region $-1.5 < \Z < -1.0$, than the other two fields (with a $\chi^{2}/\nu$ of 7.8 and 7.5 with respect to fields F1 and F2, respectively, discrepant at $\sim4.5\sigma$). The simplest explanation of this metal-poor peak, is that it is the detection of a distinct metal-poor stellar halo that becomes an increasing fraction of the total population at larger radii 

For a smooth halo, the absence of the low metallicity peak in the middle field is surprising. In Figure \ref{fig:MDF_rbins}, we further investigate NGC~3115's halo by splitting the fields into radial bins (where the distance from the center of the galaxy increases from top to bottom). These MDFs now host similar numbers of stars (resulting in similar statistics). The general trends are still observed. In particular, the low metallicity peak can still be seen in most of the inner fields despite the increased errors due to the lower number of stars in each MDF. This suggests that the absence of the low metallicity peak in the middle field is not due to the lower number of stars compared to the inner field. Instead, the non-detection of the metal-poor peak is likely due to substructure. Such substructure is expected in galactic halos because of their long relaxation times. Interestingly, the outer region of field F2 does show hints of the low metallicity population (though the reduced number of stars means that this is not significant). This suggests that we are observing some substructure in the inner regions of field F2 (at around 35~kpc). 

Over/Under densities of metal-poor halo stars can be produced if these stars were accreted into the halo from the disruption of dwarf galaxies. Many dwarf galaxies are observed to have $\Z\sim-1.3$ \citep[e.g.][]{Kirby11, Kirby13} and their accretion and stripping can result in streams of metal poor stars in the galaxy's halo. It is therefore possible that we do not detect the metal-poor peak in the middle field because it is underdense with respect to the inner and outer fields. This is consistent with Figure \ref{fig:radial_profs}, which shows that the number of metal-poor stars, and the fraction of metal-poor to -rich stars, is lower than that expected at around 35~kpc (based on a powerlaw extrapolation of the other fields). Additionally, this field may have a slight overdensity of metal-rich stars with a more peaked metal-rich population than any other field (see Figure \ref{fig:MDF_rbins}) and more metal-rich stars than expected ($\sim2\sigma$ higher, see Figure \ref{fig:radial_profs}). An over-density of metal-rich stars makes it harder to detect the minority metal-poor population by hiding it in the tail of the metal-rich distribution. Such enriched streams have been observed in other galaxy's halos, most notably the `Sagittarius stream' in the Milky Way \citep{Ibata01a} and the `giant stellar stream' in M31 \citep{Ibata01b}.

\subsection{Trends with galactocentric radius}
\label{sec:radial}

\begin{figure}
 \centering
 \vspace{-5mm}
 \includegraphics[height=88mm,angle=270]{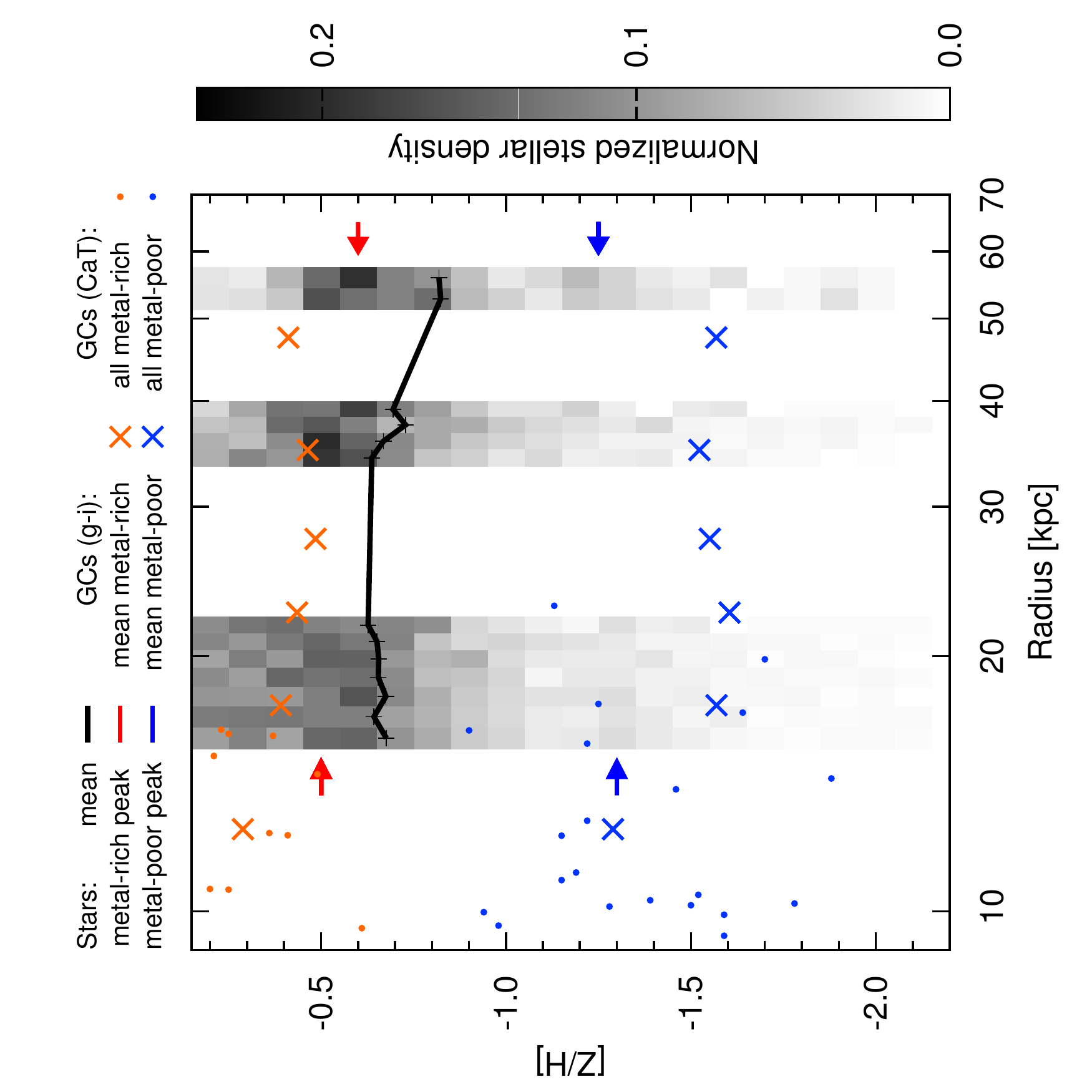} 
 \caption{MDFs as a function of radius. Each radial bin is normalized to a total of 1 over the full metallicity range investigated ($-2.25<\Z<-0.15$). Darker bins represent a higher density of stars at that radius and \Z. The MDFs are completeness corrected, but truncated for $\Z>-0.15$, where the correction becomes unreliable. The red and blue arrows indicate the locations of the metal-rich and -poor peaks in the inner and outer fields. The inner field has a broader metal-rich population and is clearly truncated at the metal-rich limit. At larger radii, the density of stars has dropped significantly before the metal-rich limit is reached. The black line tracks the mean \Z which decreases from $\Z=-0.65$, at 35~kpc, to $\Z=-0.8$, at 60~kpc. The orange and light-blue crosses show the mean metallicity of the metal-rich and -poor globular clusters, as determined from their integrated $g-i$ colors. The orange and light-blue points show individual clusters with metallicities determined from integrated spectroscopy. These globular clusters, and how they compare with NGC~3115's halo stars, are discussed in Section \ref{sec:gcs}. \\}
 \label{fig:r_MDF} 
\end{figure}

\begin{figure}
 \centering
 \includegraphics[height=86mm,angle=270]{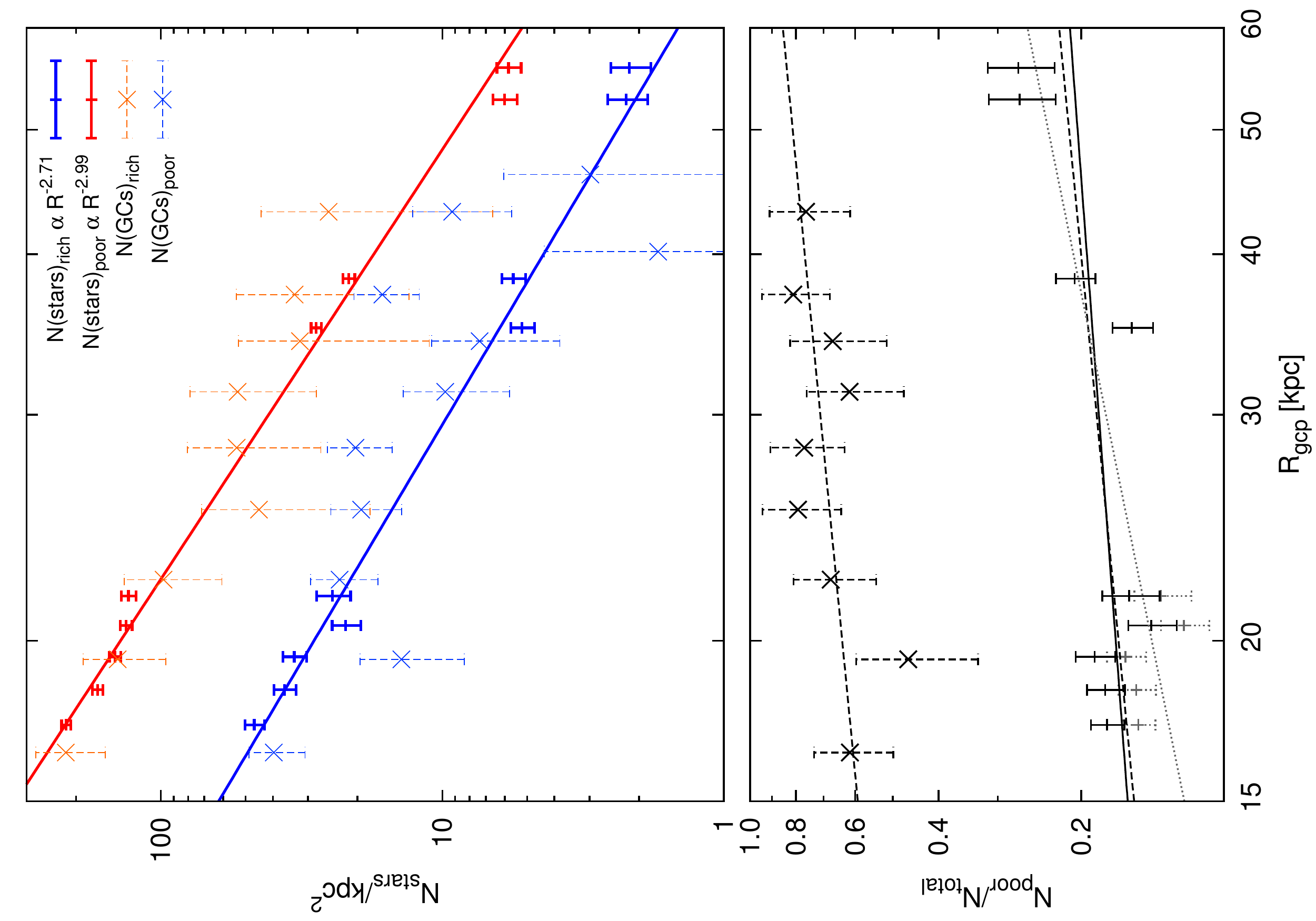} 
 \caption{{\it Top panel:} The projected density of stars and globular clusters as a function of radius from the center of NGC~3115. The dark-blue points and light-blue crosses show the metal-poor ($\Z < -0.95$) stars and globular clusters, respectively. The red points and orange crosses show the metal-rich ($\Z \geq -0.95$) stars and globular clusters, respectively. The best fitting power laws to the metal-rich and -poor stellar populations are included. The globular cluster profiles are scaled to match the density of the corresponding stellar population. {\it Bottom panel:} The fraction of metal-poor stars (points with solid errorbars) and globular clusters (crosses with dashed errorbars). The upper dashed black line is a powerlaw fit to the fraction of metal-poor globular clusters. The lower dashed black line is the same powerlaw, but scaled to fit the stellar population. The solid black line is a powerlaw fit to the fraction of metal-poor stars. The dotted gray line shows a powerlaw fit to the fraction of metal-poor stars, assuming a 20$\%$ increase in the metal-rich population in the inner field. The globular clusters results are discussed in Section \ref{sec:gcs}. \\}
 \label{fig:radial_profs} 
\end{figure}

In Figure \ref{fig:r_MDF}, we investigate the variation of metallicity as a function of radius in the halo of NGC~3115. In this figure the stellar density, as a function of radius and \Z, is indicated by the grayscale histogram, with higher density regions having darker colors. The red and blue arrows show the metal-rich and -poor peaks observed in the inner and outer MDFs. This figure highlights the trends observed in the MDFs: the fraction of stars with $\Z<-0.4$ decreases significantly from 20~kpc to 60~kpc; and the higher metallicity peak becomes slightly less enriched with increasing radius. In addition the second peak in the MDF at around $\Z=-1.3$ can be seen in many of the radial bins (with little radial variation). The black line in this plot traces the mean metallicity of the stellar population as a function of radius. This is found to decrease with increasing radius from $\Z=-0.65$ to $\Z=-0.8$. It should be noted that the mean metallicity at small radii ($\sim$20~kpc) may be slightly higher than calculated since the MDF at this radius is more truncated at our high metallicity detection limit. This may artificially produce the observed flattening of the mean metallicity between $15-35$~kpc. The orange and light-blue points and crosses in this figure show the galaxy's globular cluster population \citep[based on photometry from the catalog of][]{Jennings14}. These clusters, and how they compare with NGC~3115's stellar population, are discussed in Section \ref{sec:gcs}. 

In the top panel of Figure \ref{fig:radial_profs}, we show how the density of stars varies as a function of galactocentric radius for both the metal-rich stars ($\Z \geq -0.95$, red points) and metal-poor stars ($\Z < -0.95$, blue points). The metal-rich population, with a best fitting power law index of $-3.0$, falls off slightly more quickly than the metal-poor population, with a best fitting power law index of $-2.7$. This is consistent with a lower metallicity stellar halo population that becomes increasingly important at larger radii.  Assuming these power-laws fits, the halo will not be dominated by metal-poor stars until unrealistically large radii. However, there are suggestions that the metal-rich bulge/halo component may actually be falling off faster than this power law, with the outer field containing fewer metal-rich stars than expected. This is further demonstrated by the bottom panel of Figure \ref{fig:radial_profs}, which shows that the fraction of metal-poor stars increases significantly in the outer field. In this panel we also show the effect that truncating the MDF may have on the metal-rich population at lower radii (as gray points). These points are generated under the assumption that 20$\%$ of the metal-rich stars are above our detection limit. Assuming this correction, the data are consistent with a much steeper power law fit -- such as the dotted-gray line. Even assuming this fit, the low metallicity stellar halo will not become the dominant component until $\sim$200~kpc.

\subsection{The mass of NGC~3115's metal-poor stallar halo} 

The observed stellar density profiles, presented in Figure \ref{fig:radial_profs}, allow us to estimate the total mass of NGC~3115's metal-poor stellar halo. The data provide a good constraint on the stellar density profile for $R>15$~kpc, but poorly constrain the shape of the profile at smaller radii. Because the stellar density profile is likely to flatten at small radii, we fit it to Sersic functions, rather than the powerlaw presented in Figure \ref{fig:radial_profs}. We consider three profiles with Sersic indices of $n=2,4$ and $8$. All three profiles (and a powerlaw) are consistent with the data. For each profile, we determine the total number of metal-poor stars (with F110W~$<26.5$) by integrating over all radii. We then convert this to a total stellar mass using a luminosity function from the Dartmouth models of \citet[][assuming a 10~Gyr old stellar population with $\Z=-1.3$ and a Chabrier like IMF]{Dotter08}. The derived metal-poor halo masses are $[1.1, 1.8, 3.9] \times 10^{10}M_{\odot}$ for $n=2,4,8$, respectively. This suggests that, for reasonable stellar density profiles, our estimate of the metal-poor stellar mass is accurate to $\sim$ a factor two. 

We follow a similar method to estimate the total stellar mass of the metal-rich population. We conclude that, assuming a Sersic index of $n=4$, the total stellar masses of NGC~3115's metal-rich and -poor stellar populations are $M_{*,rich}=11.4\times10^{10}M_{\odot}$ and $M_{*,poor}=1.8\times10^{10}M_{\odot}$, respectively. This implies a stellar halo mass fraction of $\sim14\%$. Lower fractions have been estimated for both the Milky Way \citep[$\sim2\%$,][]{Bell08,Cooper13} and M~31 \citep[$\sim4\%$,][]{Courteau11}. The higher halo mass fraction estimated for NGC~3115 may be a result of differences in the evolution of early-type and late-type galaxies. The simulations of \citet{Cooper13} produce a range of stellar halo mass fractions consistent with that estimated for NGC~3115. 

We can also infer the total stellar mass of NGC~3115 from its integrated K-band luminosity ($L_{K}$). The K-band luminosity within 250$\arcsec$ ($= 12.3$~kpc) has previously been estimated from 2MASS observations to be $L_{K}=9.5\times10^{10}L_{K,\odot}$ \citep{Jarrett03}. However, our observations demonstrate a significant stellar mass beyond 12.3~kpc. We therefore extrapolate this luminosity to include the stellar mass in the galaxy's outer halo, assuming the same profiles considered above (fit to the resolved stellar population). To convert to stellar mass, we take the mass-to-light ratio to be in the range $0.78<M_{*}/L_{K}<0.87$ \citep[based on the integrated color of NGC~3115 and the relationship shown in Fig. 1 of][]{Fall13}. This implies a total stellar mass in the range $[ 9<M_{*}<17 ]\times10^{10}M_{\odot}$ -- in excellent agreement with the total mass inferred from integrating our observed stellar densities, $M_{*}=13\times10^{10}M_{\odot}$


\subsection{Summary: properties of NGC~3115's stellar halo}

Based on the MDFs presented in Figures \ref{fig:MDF_single}, \ref{fig:MDF_rbins} and \ref{fig:r_MDF}, we note the key features of NGC~3115's stellar halo: 

\begin{itemize}[leftmargin=5mm]

\item Across all three fields observed, spanning $15<R<57$~kpc ($6<R<23r_{e}$), the stellar halo is dominated by a relatively metal-rich population -- peaked in the range  $-0.6<\Z<-0.5$. 

\item The main (metal-rich) peak in the MDFs becomes slightly less enriched with increasing radius, shifting from $\Z \sim -0.45$ to $\Z \sim -0.65$ from 15~kpc to 55~kpc. Additionally, the fraction of high metallicity stars (those with $\Z>-0.4$) is much higher in the innermost field (F1) than in the two outer fields. 

\item The mean \Z drops with increasing radius from $\Z=-0.65$ at 35~kpc to $\Z=-0.8$ at 60~kpc. The inner region observed ($15-35$~kpc) has mean $\Z \sim-0.65$ and shows no significant variation. However, the MDFs of the innermost regions are clearly incomplete showing a significant fraction of stars up to our metal-rich detection limit. 

\item All three fields show a tail of low metallicity stars extending to the lowest metallicities. The fractions of metal-poor stars from $15-35$~kpc are similar, comprising 17$\%$ of the total population (although we note that the fraction may be lower in the innermost field because a larger proportion of its metal-rich stars are beyond our detection limit). The fraction of metal-poor stars in the outer field is significantly higher at $28\%$. A simple extrapolation of the population suggests that the galaxy's halo may not be dominated by metal-poor stars until $R\gtrsim200$~kpc. 

\item We observe a distinct metal-poor halo that is peaked at $\Z \sim -1.3$. This is most pronounced in the outermost field. The detection of this metal-poor peak is consistent with an underlying low metallicity stellar halo that is less concentrated than the metal-rich population. 

\item There are hints of substructure in the halo at $\sim 35$~kpc. 

\item We measure the total metal-poor stellar halo mass for NGC~3115 of $2\times10^{10}M_{\odot}$ and a total stellar mass of $13\times10^{10}M_{\odot}$. 

\end{itemize}

\section{Comparison to model MDFs} 
\label{sec:models} 

In this section, we compare the MDFs of NGC~3115's stellar halo populations to the predictions from in situ star formation using three simple chemical evolution models: a pristine closed box; a closed box with pre-enriched gas; and an accreting gas model. We take these models in the forms presented by \citet{Kirby11} and refer the reader to this paper (and the references therein) for more details. For all of these models we convert [Fe/H] to \Z assuming the $\alpha$-abundance used to derive NGC~3115's MDFs, [$\alpha$/Fe]~=~+0.4.

\subsection{Closed box model} 

In the closed box model, initially pristine gas forms stars that feed metals in to an isolated system. The gas is assumed to be well mixed at all times and no material enters or leaves the system \citep[see e.g.][]{Talbot71, Binney98}. This model predicts an MDF of the form: 

\begin{equation} 
 N \propto (10^{\rm [Fe/H]}) {\rm exp} \left( \frac{-10^{\rm [Fe/H]}}{p} \right)
\end{equation}
\\
Here $p$ is the yield of each generation of stars (a measure of the fraction of heavy elements produced). To first order, this simple prediction provides a reasonable representation of the Milky Way's bulge population \citep[e.g.][]{Rich90}. However, it predicts more metal-poor stars than are observed in the Milky Way's disk population, the long known `G-dwarf problem' \citep[as first proposed by][]{van_den_bergh62,Schmidt63} and later observed in both K-dwarfs \citep{Schlesinger12} and M-dwarfs \citep{Woolf12}. The G-dwarf problem has also been proposed based on the integrated properties of other galaxies \citep[e.g.][]{Worthey96}. 

The gray-dashed line in Figure \ref{fig:MDF_single} shows the best fitting closed box models to the three fields in the halo of NGC~3115. It can be seen that this model provides a poor fit to the data, predicting many more low metallicity stars than are observed. We thus conclude that early-type galaxy spheroids may have a similar G-dwarf problem to the Milky Way's disk.

\subsection{Pre-enriched closed box model} 

A slight adaption to the simple closed box model is to start the star formation from pre-enriched gas. The predicted MDF then takes the form: 

\begin{equation} 
 N \propto (10^{\rm [Fe/H]}-10^{{\rm [Fe/H]}_{i}}) {\rm exp} \left( \frac{-10^{\rm [Fe/H]}}{p} \right) 
\end{equation}
\\
Here, the addition of the ${{\rm [Fe/H]}_{i}}$ term accounts for the initial metallicity of the gas. This condition may be expected if stars form from gas that was enriched by earlier star formation. For example, stars in the Milky Way's disk may have formed from gas enriched by prior star formation in the Galactic bulge. By predicting less metal-poor stars, this model provides a better representation of the Milky Way's disk stars. However, the model still predicts a larger metal-poor tail than is observed. 

The best fitting pre-enriched models to NGC~3115's halo MDFs are plotted as the dark gray dot-dashed lines in Figure \ref{fig:MDF_single}. This function produces an improved fit over the pristine closed box model by reducing the number of metal-poor stars formed. However, the best fitting models to the three MDFs all require high initial metallicities of ${\rm [Fe/H]}_{i} \sim -1.8$. Even allowing for this pre-enrichment, the model still provides a poor fit, predicting a less peaked MDF than observed. 

\subsection{Accreting gas model} 

In the accreting gas model, the system is allowed to accrete pristine gas over time. The model therefore requires less initial gas (from which the lowest metallicity stars form). As the (pristine) gas is accreted, it mixes with the gas in the system (that is now enriched from earlier star formation) and hence produces higher metallicity stars. In this way, the model produces a stellar population with a lower fraction of metal-poor stars. Such a model provides a solution to the G-dwarf problem in the solar neighborhood \citep[e.g.][]{Pagel97,Prantzos98} and provides a good fit to the MDFs of the Milky Way's dwarf galaxies \citep{Kirby11}. 

We adopt the accreting gas model as presented by \citet{Kirby11} \citep[see also][for the initial discussion of this model]{Lynden-Bell75}. In this model, the gas mass ($g$) is related to the stellar mass ($s$) via: 

\begin{equation} 
g(s) = \left(1-s/M\right)\left(1+s-s/M\right)
\end{equation} 
\\
The parameter $M$ ($>1$) represents the amount of gas the galaxy accretes. The MDF of this model is given by:

\begin{equation} 
 N \propto \frac{10^{\rm [Fe/H]}}{p} \frac{
1+s(1-1/M)
}{
(1-s/M)^{-1} - 2(1-1/M) (10^{\rm [Fe/H]}/p)
}
\end{equation}

Where $s$ is given by the equation: 

\begin{multline} 
 {\rm [Fe/H]}(s) = log\left\{ 
p\left( \frac{M}{1+s-s/M} \right)^{2}  \right. \\
\left. \times \left[ ln \left(\frac{1}{1-s/M} \right) - \frac{s}{M} \left( 1-\frac{1}{M} \right)\right] \right\}
\end{multline}

We solve the equation for $s$ numerically for a given [Fe/H] and $M$ using the {\sc scipy} task {\sc root} and perform a grid search to find the best accreting gas model fit to the observed MDFs. We consider parameters in the range $0.1<M<5.0$ and $0.1<p<0.5$. Similar fits were obtained to the three observed MDFs (in different radial bins). In Figure \ref{fig:MDF_single}, we plot the best fitting accreting gas model to the inner field (F1) as the dotted black line. This model has $M=4.0$ \citep[which is in the range determined for the Milky Way's dwarfs $1.3<M<9.1$,][]{Kirby11}. This model provides a good representation of the MDF observed in field F1. The model also fits field F2 quite well, with the exception of the peak at $\Z=-0.5$. This may again be suggestive of substructure in this middle field that produces the more peaked profile observed. The outer field, F3, hosts less metal-rich stars than predicted by this model and the low metallicity peak is clearly observed above the model prediction.

\section{The outer stellar halo of NGC~3115 and other galaxies} 
\label{sec:other_gal}

Previously, \citet{Elson97} studied an HST/WFPC2 field in the halo of NGC~3115. This field is $\sim$22~kpc from the center of the galaxy, although it is closer to the major axis than our observations. Interestingly, this study also suggested a bimodal metallicity distribution in NGC~3115's halo with a metal-poor peak at [Fe/H]~$=-1.3$ (slightly more enriched than observed in our fields, assuming a similar $\alpha$-abundance). However, the implied MDF is quite different from our observations -- theirs suggested similar numbers of stars in the metal-rich and -poor populations. It was also noted by \citet{Kundu98} that this bimodal distribution can be explained as an artifact of the WFPC2 calibration, which has a break between the two proposed populations. Given the relatively low fraction of stars we observe in the low metallicity population, our results favor the interpretation of \citet{Kundu98} -- that this field is likely dominated by a single population. However, it is possible that substructure may result in this field genuinely having a larger fraction of metal-poor stars. Additional \hst observations, utilizing its updated instruments and covering a larger area, could resolve this question. 

\begin{figure}
 \centering
 \includegraphics[height=86mm,angle=270]{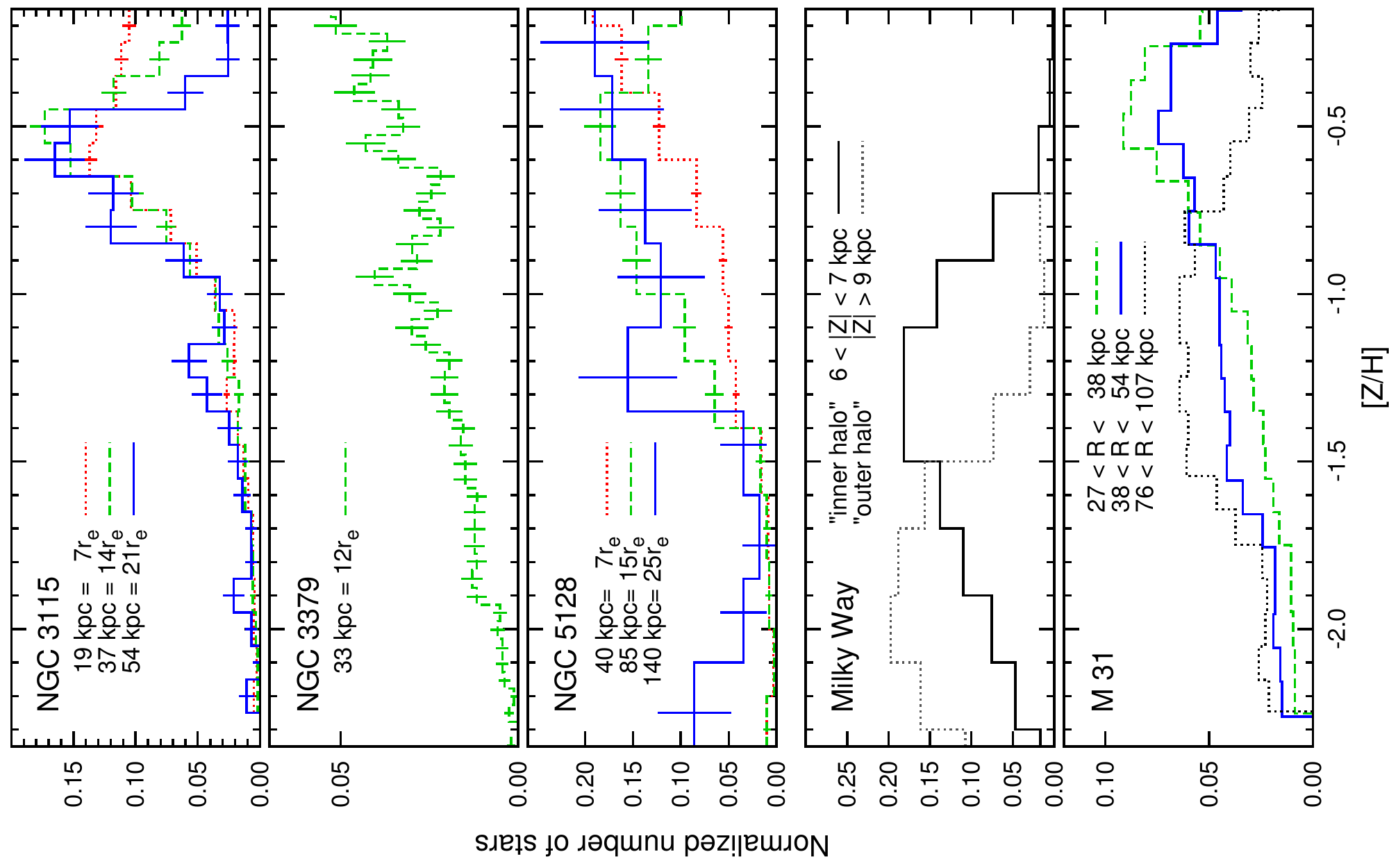} 
 \caption{The outer halo MDFs of local galaxies that have been investigated via their resolved stellar populations. The top three panels show the MDFs for the three early-type galaxies that have the \hst observations required to resolve the stellar populations in their outer halos: NGC~3115 (this study); NGC~3379; and NGC~5128. The dotted-red, dashed-green and solid-blue lines show the MDFs at similar effective radii. The bottom two panels show the MDFs for two late-type galaxies, the Milky Way and M31. For M31, the dashed-green and solid-blue lines show MDFs at similar physical radii to those in NGC~3115. \\}
 \label{fig:compare_mdfs} 
\end{figure}

Figure \ref{fig:compare_mdfs} compares NGC~3115's halo with that of other galaxies. Only two other early-type galaxies have \hst observations of their halos at similarly large $r_{e}$. An ACS field 12$r_{e}$ from the center of NGC~3379 was observed by \citet{Harris07a}. The derived MDF is shown in the second panel of Figure \ref{fig:compare_mdfs}. A population of low metallicity stars is observed, but this is part of a broader MDF than observed in NGC~3115. This observation was at similar radii to our middle field. The only early-type galaxy that has been observed at similarly large effective radii to our outer field (where the low metallicity halo is likely to be most significant) is NGC~5128. Eight fields, four WFC3/UVIS + four ACS/WFC (taken in parallel at a similar location), were observed in NGC~5128's halo from 10$r_{e}$ out to 25$r_{e}$ \citep{Rejkuba14}. In the third panel of Figure \ref{fig:compare_mdfs}, we show the derived MDFs for three halo fields at similar $r_{e}$ to our observations of NGC~3115. It can be seen that NGC~5128 has a broadly similar halo to that observed in NGC~3115, with: the stellar population peaked at relatively high metallicity; the median metallicity reducing with increasing radius; and the fraction of metal-poor stars increasing in the outer field. In the outermost field of NGC~5128 there are also hints of the metal-poor peak that is observed at $\Z=-1.3$ in NGC~3115's halo. However, the number of stars in this outer field (58) is significantly lower than that of our NGC~3115 analysis. This limits our ability to significantly test for this metal-poor population. 

The properties of early-type galaxy halos, as inferred from integrated surface brightness photometry or spectroscopy are normally limited to more central regions than probed by this study. However, several studies have pushed galaxy color profiles to quite large effective radii. For example, \citet{LaBarbera12} used the combined profiles of 674 early-type galaxies in the Sloan Digital Sky Survey (SDSS) to show clear color gradients out to 8$r_{e}$. This color profile suggests that by 8$r_{e}$ the galaxies have mean metallicities in the range $-1.0<\Z<-0.7$. This range is consistent with the the mean metallicity of NGC~3115's halo (see the black solid line in Figure \ref{fig:radial_profs}). Recently, \citet{Mihos13} used integrated photometry to trace (the early-type galaxy) NGC~4472's halo out to $7r_{e}$ (100~kpc). Interestingly, they identified a color gradient which, if purely due to metallicity variations, would imply a halo dominated by metal-poor stars with $\Z < -1.0$. This halo would be quite unlike the one we observe in NGC~3115 which is dominated by significantly more enriched stars out to much larger $r_{e}$. It is interesting that these studies suggest that NGC~3115's halo is more enriched than these more massive galaxies at similar $r_{e}$. However, caution should be used when comparing galaxy halos based on similar effective radii, since these can correspond to very different physical radii. Future observations may be able to push to fainter surface brightness limits around more early type galaxies \citep[by using, for example, the dedicated Dragonfly telescope;][]{Abraham14}. Such observations can cover a large fraction of a galaxy's halo and detect the stellar halo around relatively distant galaxies. They will therefore provide important complimentary data to \hst observations of the resolved stellar population. 

The fourth panel of Figure \ref{fig:compare_mdfs} shows the Milky Way's `inner' (${\rm 6<|Z|<7~kpc}$) and `outer' (${\rm |Z|>9~kpc}$) halo MDFs as measured by \citet[][and converted to \lbrack Z/H\rbrack \ assuming \lbrack$\alpha$/Fe\rbrack \ = \ +0.3]{Carollo10}. It can be seen that NGC~3115's halo is markedly distinct from the Milky Ways (which is dominated by metal-poor stars at similar radii). However, caution should be used when directly comparing the halo of NGC~3115 with the Milky Way (a late-type galaxy with little evidence of a bulge contribution to its halo). It should also be noted that the Milky Way's halo may not be typical of all spiral galaxies. Indeed, the presence of a low metallicity population in M31's halo was only clearly identified with resolved stellar photometry \citep{Ibata14} and spectroscopy \citep{Gilbert14} of a large fraction of its halo to beyond 100~kpc. 

In the bottom panel of Figure \ref{fig:compare_mdfs}, we also show the MDFs of M31's halo within three radial bins \citep[taken from][]{Ibata14}. Over the range $20-100$~kpc the halo MDF shifts to lower metallicities. However, the lower metallicity population only becomes dominant at much larger radii than in the Milky Way's halo -- beyond $\sim$70~kpc. Additionally, some regions of M31's outer halo are more enriched than the MDFs shown \citep[which exclude regions of spatially distinct higher metallicity stars, most notably the `giant stellar stream',][]{Ibata01b,Ibata14}. At a similar physical radius to the outer observation of NGC~3115's halo, M~31's halo has a larger fraction of metal-poor stars, but is otherwise remarkably similar (with metal-rich and metal-poor populations peaked at \Z~$\sim -0.6$ and \Z~$\sim -1.3$). We again note that caution should be used when comparing early- and late-type galaxies, but it can be seen that M~31's halo is not dominated by metal-poor stars until $\gtrsim70$~kpc, which is more distant than we observe in NGC~3115.

\section{Tracing stellar halos with globular clusters}
\label{sec:gcs}

\begin{figure}
 \centering
 \includegraphics[height=86mm,angle=270]{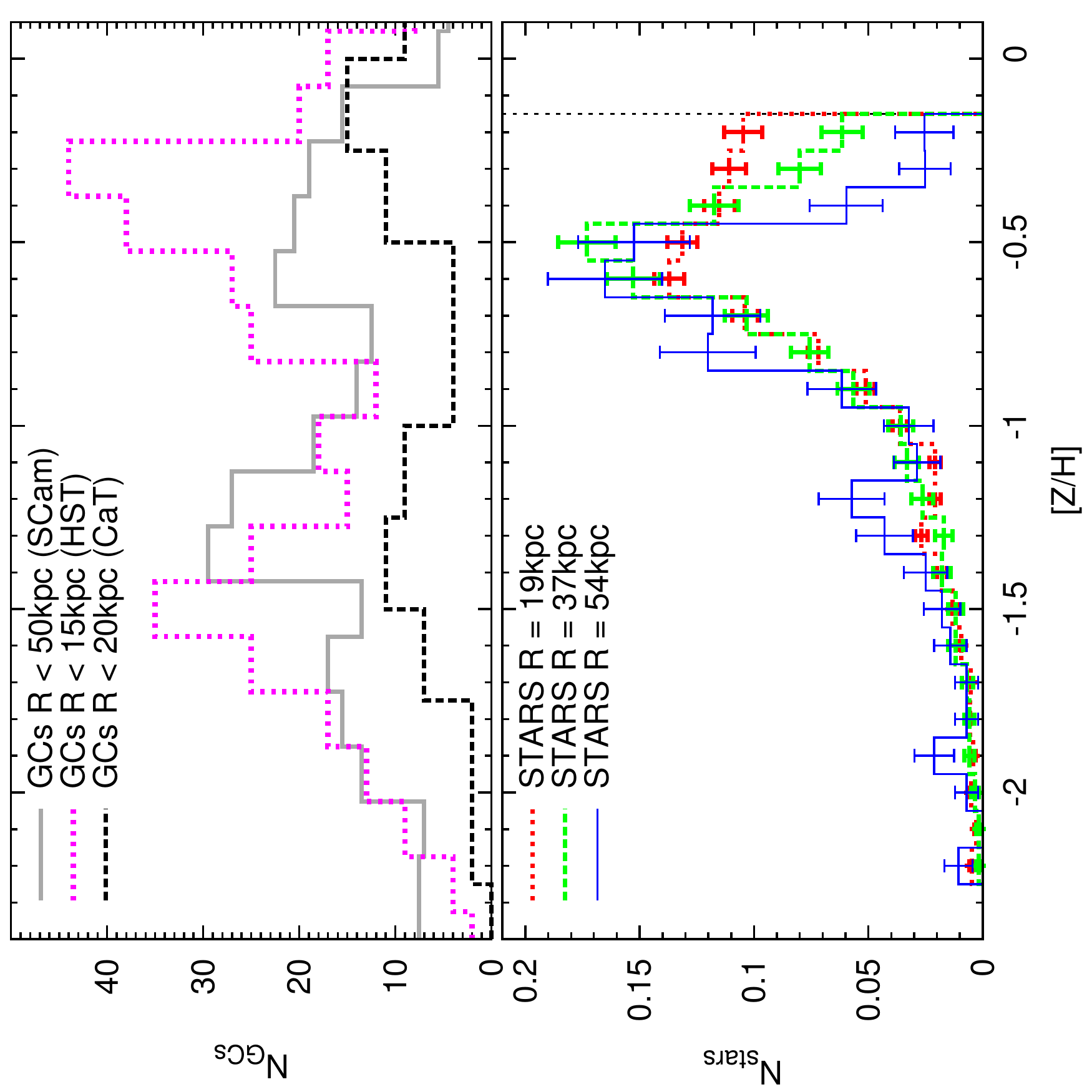} 
 \caption{The MDFs of NGC~3115's globular cluster population (top) and stellar halo population (bottom). The different lines in the globular cluster MDFs show \Z as inferred from: spectroscopy of a sample of clusters out to 20~kpc (dashed-black); $g-z$ color from \hst observations out to 15~kpc (dotted-pink); and $g-i$ color from ground based observations out to 50~kpc (scaled to ${\rm N_{GCs}}/2$, solid-gray). The stellar MDFs are as in the previous Figures and are centered at: 19~kpc (dotted-red); 37~kpc (dashed-green); and 54~kpc (solid-blue). \\}
 \label{fig:mdf_gcs} 
\end{figure}

As bright chemo-dynamical tracers of the stellar population of a galaxy, globular clusters are thought to be powerful probes of the outer halos of galaxies (where their stellar populations are often challenging/impossible to observe directly). The observations presented here allow us to directly compare the stellar halo of NGC~3115 with inferences from the galaxy's globular cluster system. 

The globular cluster systems of early-type galaxies have two key features that provide clues to the stellar population of their halos. Firstly, it is well established that the globular cluster populations of most galaxies have a bimodal metallicity distribution with a metal-rich population peaking at $\Z \sim -0.5$ and a metal-poor population peaking at $\Z \sim -1.5$ \citep[e.g.][]{Peng06,Strader06}. Secondly, it has been observed in many galaxies that the metal-poor clusters are more spatially extended than the metal-rich clusters \citep[e.g.][]{Bassino06,Brodie06,Forbes11}. An interpretation of these two common properties of globular cluster systems is that most galaxies have a distinct, extended, low metallicity stellar halo that extends to many effective radii and is associated with the metal-poor globular cluster population. However, this interpretation is poorly tested, particularly in early-type galaxies. 

NGC~3115 has a large globular cluster system with a strongly bimodal MDF \citep[e.g.][]{Arnold11, Brodie12, Cantiello14,Jennings14}. The metal-poor clusters are observed to be less centrally concentrated than the metal-rich clusters, with the ratio of low- to high-metallicity clusters increasing with radius \citep[e.g.][]{Arnold11}. In the top panel of Figure \ref{fig:mdf_gcs}, we plot the MDFs of NGC~3115's globular cluster system based on \hst observations and the suite of data available from the SLUGGS survey \citep{Brodie14}. The black-dashed line shows the globular cluster MDFs from spectroscopic observations utilizing the CaT index \citep[sensitive to the Ca~{\sc ii} triplet;][]{Brodie12}. These data cover a sample of clusters out to a radius of 20~kpc. We also consider the MDFs based on the recent globular cluster photometric catalog of \citet{Jennings14}. The purple-dotted line shows the MDF based on \hst observations of the inner region of the galaxy ($\lesssim 15$~kpc), converted from $g-z$ color to \Z based on the empirical relationship of \citet{Peng06}. The gray-solid line shows the MDF based on $g-i$ color from Subaru/Suprime Cam observations, converted to \Z based on the conversion presented in \citet{Usher12}. These data cover clusters out to larger radii ($\lesssim 50$~kpc). However, without high resolution imaging or spectroscopic confirmation, there is likely to be more contamination from non-cluster sources in these data. 

It can be seen that all measurements show the globular cluster system to be strongly bimodal. However, the peak of the metal-rich and metal-poor populations varies slightly between the different measurements with peaks in the range $-0.5 < \Z < -0.2$ and $-1.5 < \Z < -1.3$, respectively. Some of this variation may be attributed to radial variations, since the globular cluster peaks may become bluer with increasing radius \citep[see Figure \ref{fig:r_MDF} and][]{Jennings14}. However, variation may also be due to systematic errors in the methods used to convert color and spectroscopic index to \Z. This is a particular issue for the metal-poor clusters where metallicity variations result in only small color variations. We adopt a metallicity of $\Z=-0.7$ as the boundary between the metal-rich and -poor globular clusters. This is similar to the boundary adopted by \citet{Jennings14} and provides a good split between the two peaks observed in the globular cluster MDFs (see Figure \ref{fig:mdf_gcs}). However, this is slightly more enriched than the boundary chosen for the stellar population ($\Z=-0.95$). We choose to keep these different boundaries and note that some of this offset may result from the relatively small fraction of metal-poor stars. The effect of this is that the MDF is dominated by the tail of the metal-rich population to relatively low metallicities. 

The bottom panel of Figure \ref{fig:mdf_gcs} shows the MDF of the three stellar halo fields presented in this paper. The MDFs of these fields have metal-rich peaks in the range $-0.6 \lesssim \Z \lesssim -0.5$. This is similar to the metal-rich globular cluster population, though perhaps offset to lower metallicities by 0.1-0.2~dex. Some of this shift may result from the larger galactocentric radii of the stellar halo populations relative to the majority of these globular clusters. Indeed, it can be seen from Figure \ref{fig:r_MDF} that, in a similar fashion to the stellar population, the metal-rich globular clusters become less enriched with increasing radius (as inferred from their integrated $g-i$ colors). Based on comparing the integrated colors of galaxy bulges with their globular cluster colors, \citet{Spitler10} have previously proposed that the metal-rich globular clusters may be slightly less enriched than the stellar bulge stars of massive galaxies. For a galaxy with the mass of NGC~3115\footnote{$M=8.2\times10^{10}M_{\odot}$, based on its K-band luminosity \citep[$L_{K}=9.5\times10^{10}L_{K,\odot}$][]{Jarrett03} and assuming  $M/L_{K}=0.86$ \citep[as used by][]{Spitler10} }, their work suggests that the metal-rich globular clusters may be less enriched by $\sim$0.2~dex. Our data are consistent with such a shift. However, we also note that this offset could be produced by systematic uncertainties in the different methods used to estimate metallicity (stellar CMD fitting for the stars relative to stellar population fitting to integrated colors/spectral indices for the globular clusters). This is evident from top panel of Figure \ref{fig:mdf_gcs}, where similar sized offsets can be seen between the metal-rich peaks of the globular cluster population derived from different methods. Further investigation of the metallicities derived from these different methods is clearly important, but beyond the scope of this paper. 

Figure \ref{fig:mdf_gcs} also shows that the second, lower metallicity, peak observed in the stellar halo is at a similar metallicity to the peak of the metal-poor globular clusters. This peak is likely the detection of the metal-poor stellar halo of NGC~3115, which is associated with its metal-poor globular clusters. We note that, based on their $g-i$ colors, the metal-poor globular clusters may shift to lower metallicities at larger radii (beyond 15~kpc). This can be seen in Figure \ref{fig:r_MDF}, where the mean metallicity of the metal-poor globular clusters decreases from $\Z\sim-1.3$ at small radii to a constant $\Z$ of $\sim-1.6$ from $20-50$~kpc. The peak of the metal-poor stellar halo shows a similar lack of radial variation over $20-50$~kpc, but is more enriched by $\sim$0.3~dex. We again note that this shift might be explained by errors in the conversion of stellar CMDs and integrated color to metallicity. This is a particular issue at these lower metallicities, where variations in metallicity result in relatively small variations in color. 

In Figure \ref{fig:radial_profs}, we show the radial profiles of NGC~3115's metal-rich and -poor globular clusters (as orange and light blue crosses, respectively) and stars (as red and blue points, respectively). These globular cluster profiles were produced using the Subaru/Suprime Cam photometry, published in the catalog of \citet{Jennings14}. It is likely that a significant fraction of these globular cluster candidates in the outer regions are misclassified background galaxies. We estimated this background for both the metal-rich and -poor clusters based on the density of sources in the outermost regions ($\sim60$~kpc). The globular cluster profiles were then scaled to fit the number of stars in the corresponding stellar population. Due to the lower number of sources, the globular cluster profiles have significantly larger errors than the stellar profiles. However, it can be seen that the radial profile of NGC~3115's globular clusters is consistent with the galaxy's halo stars. Specifically, it can be seen that metal-rich globular clusters have a steeper profile than the metal-poor globular clusters, as observed in the stellar populations. This is highlighted by the bottom panel of Figure \ref{fig:radial_profs}, where the fractions of metal-poor globular clusters and stars are observed to increase with increasing radius. 

The major difference observed between the stellar and globular cluster populations is that the fraction of metal-poor stars in the halo of NGC~3115 is much smaller than the fraction of metal-poor globular clusters. This can be seen in the bottom panel of Figure \ref{fig:radial_profs}, which shows the fraction of metal-poor globular clusters and stars. For both populations, this fraction increases with radius. However, the fractions are very different. A similar difference is observed in the Milky Way, where 70$\%$ of its globular clusters are associated with its metal-poor halo \citep[e.g.][]{Bica06}, but the Milky Way's metal-poor halo comprises only $\sim$2$\%$ of its total stellar population \citep{Carollo10}. The fraction of metal-poor globular clusters is also observed to be much higher than the fraction of metal-poor stars in the outer halo ($>8$~kpc) of NGC~5128 \citep[][]{Harris02}. \citet{Harris02} also calculate the specific frequency of NGC~5128's globular clusters ($S_{N}$, a measure of the number of GCs to a galaxy's V-band luminosity) as a function of metallicity. They find that $S_{N}$ increases with decreasing metallicity. From the stellar and globular cluster density profiles presented in Figure \ref{fig:radial_profs}, we calculate that the specific frequency of the metal-poor population in NGC~3115's halo is 7.5 times higher than that of the metal-rich population. The reason for this difference is currently uncertain. It may suggest that, in metal-poor environments, relatively more stars form in larger, denser clusters. Or it may be that clusters experience less destruction in these metal-poor environments (perhaps because they originate in dwarf galaxies that then accrete into the halos of these larger mass galaxies).

Our observations of NGC~3115's stellar and globular cluster populations present a consistent picture of the galaxy's halo. They both suggest a metal-poor stellar population, peaked at $\Z \sim -1.3$, that is more spatially extended than a more enriched population, with $\Z \sim -0.5$. We note that the major difference between the stellar and globular cluster populations is in the ratio of the metal-poor to -rich stars/ clusters, which is much smaller for the stellar population. 

The MDF and radial profile of NGC~3115's globular cluster population is consistent with its observed stellar halo. This supports the use of globular clusters as tracers of the stellar populations of early-type galaxies. Our conclusions are consistent with the findings of other studies that have shown that, out to a few $r_{e}$, the metal-rich globular clusters have similar metallicities to those of early-type galaxies stellar populations, as inferred from their integrated colors \citep{Spitler10} and spectra (Pastorello, Forbes, et al., in prep.).

\section{Conclusions}

We use deep \hst photometry to study the outer halo of the early-type galaxy NGC~3115. These observations detect and resolve thousands of the galaxy's RGB stars, two magnitudes beneath the TRGB. The TRGB is obvious in these data and we measure a TRGB distance modulus for NGC~3115 of $30.05\pm0.05\pm0.10_{sys}$ -- where the calibration of the TRGB magnitude through the F110W filter dominates the uncertainty. 

By comparing to the Dartmouth stellar isochrones, we produce MDFs of NGC~3115's halo over the range $15-60$~kpc ($6-23r_{e}$). We show that, even at these large galactocentric radii, the majority of stars are quite enriched, with peaks in the MDFs shifting from $\Z \sim -0.5$ at 15~kpc to $\Z \sim -0.65$ at 60~kpc.

All three fields have a tail of stars extending to low metallicities. We find that the mean metallicity and the ratio of metal-poor to -rich stars increases significantly with galactocentric radius. We detect a distinct metal-poor population 4$\sigma$ above the low metallicity tail in the inner and outer fields. This is apparent as a second peak in the MDF at $\Z \sim -1.3$. The reason for the absence of this peak in the middle field is uncertain, but we suggest that substructure is the likely cause. This second peak is consistent with hierarchical halo formation theories which predict a relatively low concentration metal-poor halo that becomes increasingly significant at larger radii. By extrapolating the stellar density profiles observed, we estimate the total stellar mass of NGC~3115's metal-poor halo to be in the range $[1<M_{*}<4]\times10^{10}M_{\odot}$ ($\sim14\%$ of the galaxy's total stellar mass). 

The stellar halo compares well with the expectations from NGC~3115's globular cluster population. The densities of halo stars and globular clusters appear to decrease in a similar fashion with increasing radius, such that the metal-poor population has a shallower profile than the metal-rich population. The low metallicity peak observed at $\Z \sim -1.3$ in the stellar population is likely associated with the metal-poor globular cluster population. This is the strongest evidence to date that globular clusters trace the stellar populations in the halos of early-type galaxies. However, the fraction of metal-poor to -rich globular clusters is much higher than that of the stars. This is similar to the difference observed in the Milky Way. The relatively large number of globular clusters associated with the metal-poor stellar halo, and the fact that the globular clusters are found to have a similar metallicity and profile to the metal-poor stars, supports their use as bright chemo-dynamical tracers of the stellar halos of more distant galaxies.  

Future \hst observations of more fields in NGC~3115's outer halo will be important in improving our knowledge. By covering more area, we can test for substructure in the halo and detect more stars in the low density outer regions, thereby increasing the significance of our conclusions. Of particular importance will be more distant observations in NGC~3115's halo, beyond 100~kpc, where the metal-poor halo is projected to be more dominant. In addition, integrated surface brightness photometry of the stellar halo of NGC~3115 at large radii would be powerful in comparing with our stellar population analysis. Such observations would allow us to test for substructure in NGC~3115's halo.

\section*{Acknowledgments}

This work is supported in part by HST-GO-13048.01, HST-GO-13048.002-A and by National Science Foundation grant AST-1211995. This research has made use of NASA's Astrophysics Data System. We also made use of the up to date python/pyraf/stsdas tasks installed using the Ureka package (http://ssb.stsci.edu/ureka/). MBP thanks the STSci Ureka helpdesk for assistance with an issue running the tweakshift task installed by this package on Mac OSX.

\bibliographystyle{apj_w_etal}
\bibliography{bibliography_etal}

\newpage

\begin{appendix}

\section{The effect of stellar luminosity range on the derived MDFs} 

\begin{figure*}
  \centering
  \begin{tabular}{cc}
    \includegraphics[width=80mm,angle=0]{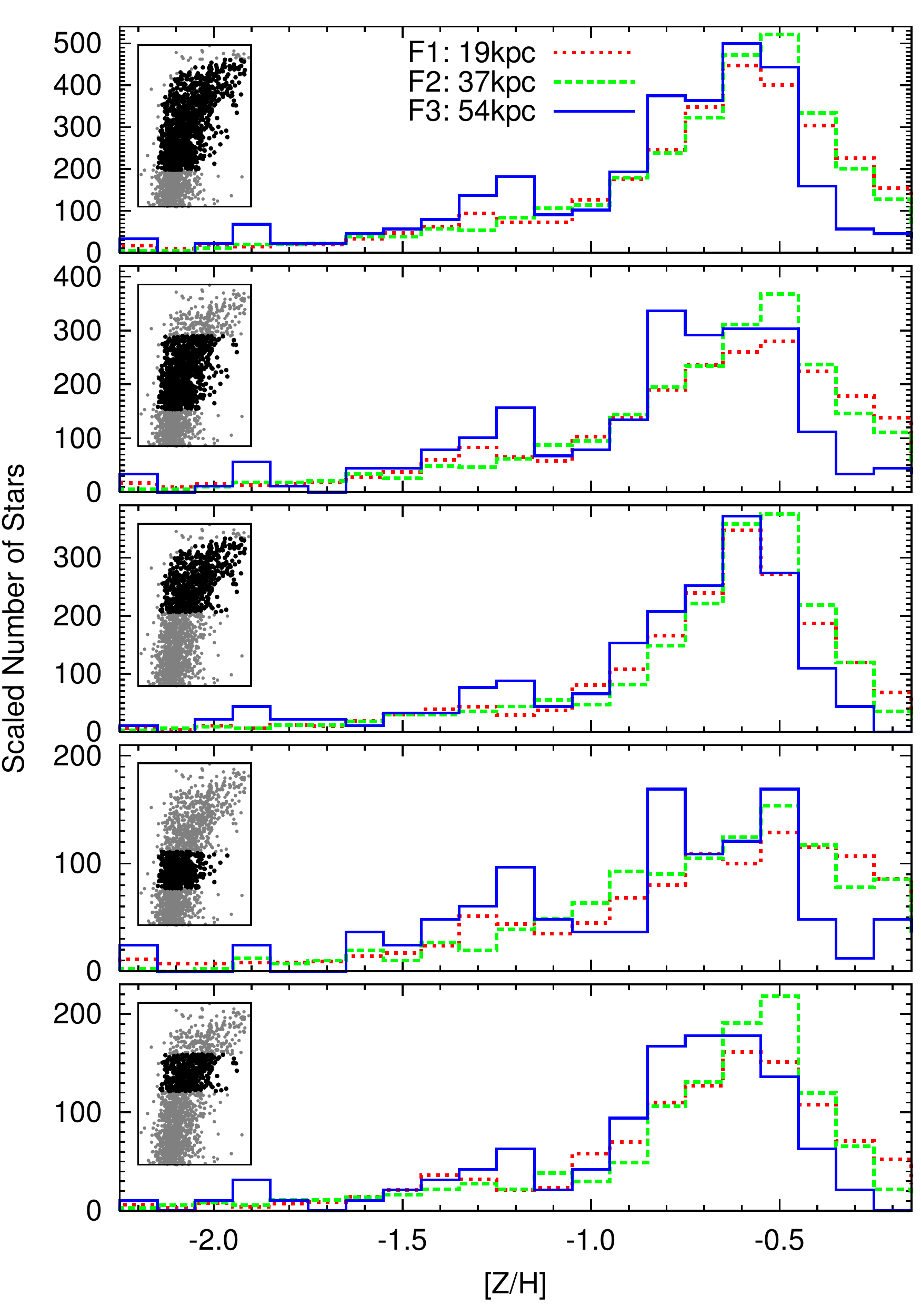} &
    \includegraphics[width=80mm,angle=0]{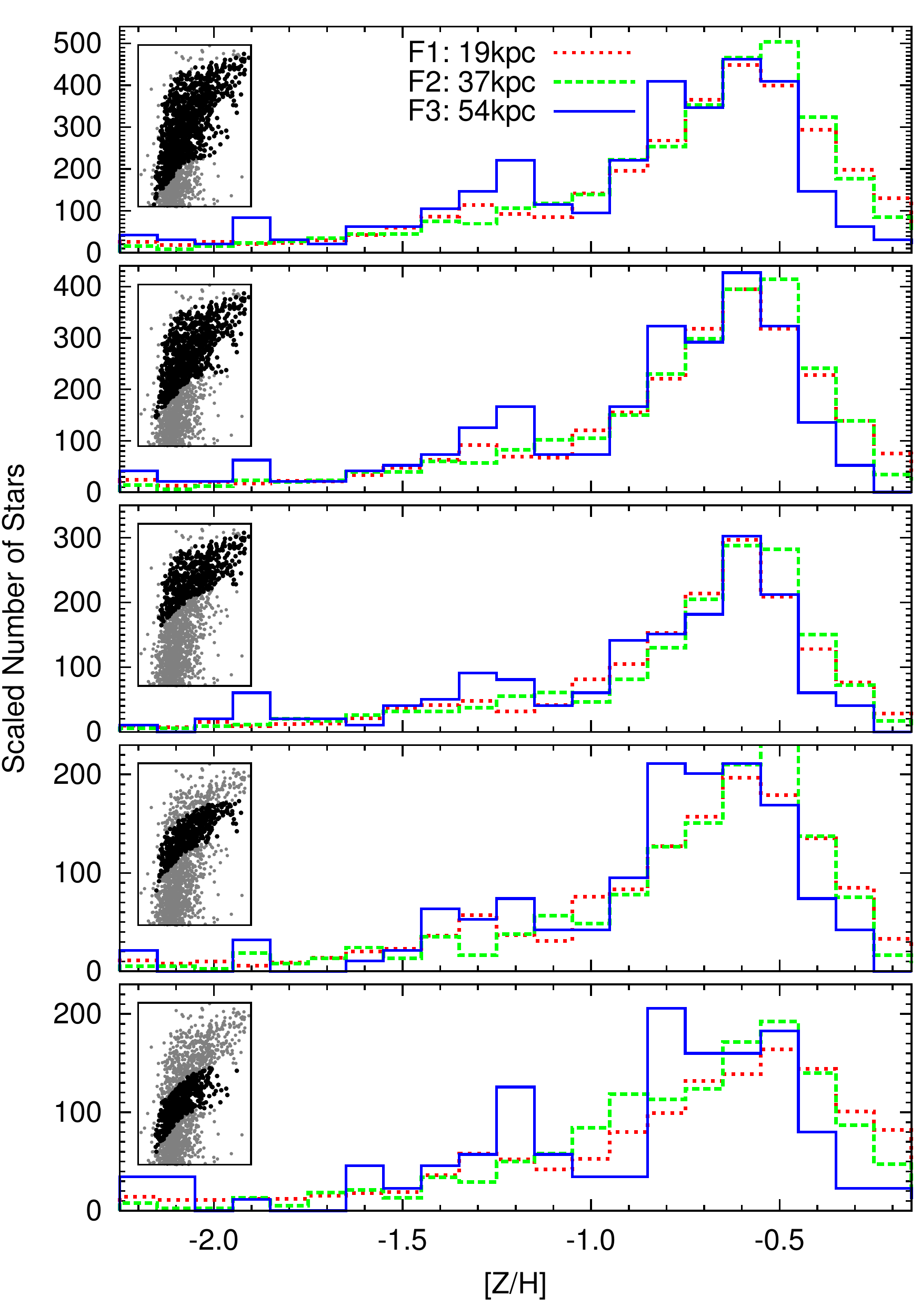} 
  \end{tabular}
  \caption{The metallicity distributions of stars in fields F1 (red-dotted), F2 (green-dashed) and F3 (blue-solid), scaled to have the same total number of stars as F1. The different panels adopt different magnitude cuts for stars used to construct the histograms. The left panels consider stars within F110W magnitude cuts. The right panels consider stars with magnitude cuts relative to the TRGB for their measured metallicity. The inserts are CMDs for the F2 field (gray) with the stars included in the MDF highlighted in black. It can be seen that the choice of magnitude cut has little effect on the characteristics of the MDFs in each field. We restrict our analysis of the MDFs to $\Z < -0.15$ -- at higher metallicities, the MDFs become increasingly unreliable due to incompleteness. }
  \label{fig:MDF_multi}
\end{figure*}

To construct MDFs of the observed stellar populations, we must choose which regions of the CMDs to study. This choice is non-trivial. We wish to include as many stars as possible while only including those stars with reliable color-magnitude (and hence metallicity) information. In Figure \ref{fig:MDF_multi}, we investigate how the choice of different magnitude limits effects the resulting MDFs. These cuts go down to F110W=26.5, where the average error on the F606W-F110W color is $\lesssim$0.15. The region of the CMD studied is indicated in the insert of each panel. The left panels consider simple cuts based on the F110W magnitude of the stars. The right panel takes F110W magnitude cuts relative to the TRGB magnitude, noting that this varies as a function of metallicity. It can be seen that, while some variation is observed due to the cuts chosen, the general features of the MDFs -- i.e. the location of the primary and secondary peaks -- are quite robust to this choice. 

\end{appendix}

\label{lastpage}

\end{document}